\documentclass[fleqn,usenatbib]{mnras}

\usepackage[T1]{fontenc}

\DeclareRobustCommand{\VAN}[3]{#2}
\let\VANthebibliography\thebibliography
\def\thebibliography{\DeclareRobustCommand{\VAN}[3]{##3}\VANthebibliography}

\usepackage{graphicx}	
\usepackage{amsmath}	
\usepackage{booktabs, caption}
\usepackage[flushleft]{threeparttable}
\PassOptionsToPackage{hyphens}{url}\usepackage{hyperref}
\usepackage{multirow}
\usepackage{times}
\usepackage[hyphenbreaks]{breakurl}
\usepackage{bm}
\usepackage{lastpage}
\newcommand{\degg}{\hbox{$^\circ$}}

\newcommand{\xmm}{{\it XMM-Newton}}

\newcommand{\arcs}{\hbox{$^{\prime\prime}$}}

\newcommand{\ls}
{\mathrel{\hbox{\rlap{\hbox{\lower4pt\hbox{$\sim$}}}\hbox{$<$}}}}
\newcommand{\gs}
{\mathrel{\hbox{\rlap{\hbox{\lower4pt\hbox{$\sim$}}}\hbox{$>$}}}}

\title[The flaring corona in PDS 456]
  {The flaring X-ray corona in the quasar PDS 456}
\author[J.N. Reeves]
  {J.N.~Reeves$^{1,2}$\thanks{e-mail: \href{mailto:james.n.reeves456@gmail.com}{james.n.reeves456@gmail.com}}, V.~Braito$^{2,1}$, D.~Porquet$^{3}$, 
  A.P.~Lobban$^{4}$, G.A.~Matzeu$^{4,5}$, E.~Nardini$^{6,7}$ \\
  $^1$Department of Physics, Institute for Astrophysics and Computational Sciences, The Catholic University of America, Washington, DC 20064, USA \\
  $^2$INAF, Osservatorio Astronomico di Brera, Via Bianchi 46 I-23807 Merate (LC), Italy \\
  $^3$Aix-Marseille Univ., CNRS, CNES, LAM, Marseille, France \\
    $^4$European Space Agency (ESA), European Space Astronomy Centre (ESAC), E-28691 Villanueva de al Ca\~{n}ada, Madrid, Spain \\
    $^5$Department of Physics and Astronomy (DIFA), University of Bologna, Via Gobetti 93/2 -- 40129 Bologna, Italy \\
  $^6$INAF, Osservatorio Astrofisico di Arcetri, Largo Enrico Fermi 5, I-50125 Firenze, Italy \\
  $^7$Dipartimento di Fisica e Astronomia, Universit\`a di Firenze, via G. Sansone 1, I-50019 Sesto Fiorentino, Firenze, Italy \\ }
\date{\today; submitted to MNRAS}

\pagerange{\pageref{firstpage}--\pageref{lastpage}} \pubyear{2020}
\begin{document}
\label{firstpage}
\pagerange{\pageref{firstpage}--\pageref{lastpage}}
\maketitle

\begin{abstract}

New {\it Swift} monitoring observations of the variable, radio-quiet quasar, PDS\,456, are presented. A bright X-ray flare was captured in September 2018, 
the flux increasing by a factor of 4 and with a doubling time-scale of 2 days. 
From the light crossing argument, the coronal size is inferred to be $\ls30$ gravitational radii for a black hole mass of $10^{9}$\,M$_{\odot}$
and the total flare energy exceeds $10^{51}$\,erg. A hardening of the X-ray emission accompanied the flare, with the photon index decreasing from 
$\Gamma=2.2$ to $\Gamma=1.7$ and back again. 
The flare is produced in the X-ray corona, the lack of any optical or UV variability being consistent with a constant accretion rate.
Simultaneous {\it XMM-Newton} and {\it NuSTAR} observations were performed, $1-3$ days after the flare peak and during the decline phase.
These caught PDS\,456 in a bright, bare state, where no disc wind absorption features are apparent. The hard X-ray spectrum shows a high energy roll-over, with an 
e-folding energy of $E_{\rm fold}=51^{+11}_{-8}$\,keV. 
The deduced coronal temperature, of $kT=13$\,keV, is one of the coolest measured in any AGN and PDS\,456 lies well below the predicted pair annihilation line in X-ray corona.
The spectral variability, becoming softer when fainter following the flare, is consistent with models of cooling X-ray coronae.
Alternatively, an increase in a non-thermal component could contribute towards the hard X-ray flare spectrum. 

\end{abstract}


\section{Introduction}

Active Galactic Nuclei (AGN) are powered by accretion on to a Super Massive Black Hole (SMBH), as a result of 
viscous dissipation of gravitational potential energy in an accretion disc \citep{Shakura73}. 
For the typical mass of SMBHs in AGN, which range from $10^6$\,M$_{\odot}$ to $\gs10^{9}$\,M$_{\odot}$, most of the accretion power is radiated in the UV band, 
in the form of the so-called Big Blue Bump \citep{Malkan82, Czerny87}. The disc emission is generally too cool to emit significant X-ray radiation, which is instead produced via Comptonization; i.e. through inverse Compton scattering of disc photons in an X-ray corona of hot electrons close to the black hole 
\citep{Haardt91,Haardt93}. Early high energy observations of Seyfert 1 galaxies, e.g. with the {\it OSSE} instrument on-board the Compton Gamma-Ray Observatory (CGRO), showed that the power-law X-ray continuum rolls over at high energies above 100\,keV \citep{ZLM93, Madejski95, Gondek96}, implying that the X-rays arise predominately via thermal Comptonization. 

A variety of X-ray observatories have since demonstrated the almost ubiquitous presence of high energy cut-offs in the hard X-ray spectra of AGN; e.g. 
{\it Beppo-SAX} \citep{Petrucci01, Perola02, Dadina07}, {\it Integral} \citep{Bassani06, Molina09, Malizia14}, the Neil Gehrels Swift Observatory (hereafter {\it Swift}) Burst Alert Telescope (BAT) \citep{Ricci18} and recently {\it NuSTAR} \citep{Fabian15, Fabian17, Tortosa18, Kamraj18, Middei19}. 
The high energy cut-off usually occurs at energies of around 100\,keV or higher, where 
the coronal electron temperatures are typically inferred to lie in the range $kT\sim50-200$\,keV (e.g. \citealt{Malizia14, Ricci18}). 

The increase in sensitivity afforded by the imaging hard X-ray optics on-board {\it NuSTAR} \citep{Harrison13} 
have made it possible to extend these coronal studies to AGN with lower X-ray fluxes, when compared to the typical, nearby, X-ray bright Seyfert 1s, 
thereby expanding the parameter space of observations.  
As a result, a small number of AGN are now thought to show high energy cut-offs at lower energies than previously thought and where the coronal temperatures 
may be as low as $kT\sim10-30$\,keV \citep{Balokovic15, Tortosa17, Kara17, Turner18}. Indeed, from studying the large number of AGN available from the BAT hard X-ray survey, \citet{Ricci18} deduced a possible anti-correlation between the cut-off energy and Eddington ratio, whereby the coronal temperature decreases with increasing ratio. Thus it is interesting to exploit new observations to test whether the typical X-ray coronal properties may differ in the high Eddington regime.

It is also well established that the X-ray emission from AGN is rapidly variable, on time-scales as short as a few kiloseconds (e.g. \citealt{MDP93}). 
Measurements of X-ray variability, characterised via the break frequencies in power spectra and from inter-band X-ray lags, are consistent with the 
corona extending down to a few gravitational radii from the black hole and with time-scales which appear to scale with the SMBH mass 
(e.g. \citealt{Markowitz03, McHardy06, DeMarco13, Kara16}). This requires that coronae are compact and radiatively efficient, where the analysis of the coronal compactness versus their temperature reveal that most AGN lie close to the relations predicted for pair production and annihilation in the corona 
(e.g. see \citealt{Fabian15} and references therein). Pair-production 
then provides a necessary mechanism to regulate the coronal temperature; here if the temperature becomes too high an excess of pairs will be produced, 
leading to a redistribution of particle energies and subsequently quenching the X-ray luminosity. The corona is likely powered by magnetic reconnection events 
(e.g. \citealt{DiMatteo98}), as the thermal energy stored in the corona itself is insufficient \citep{Fabian17} and such events may account for the rapid X-ray flares 
observed towards many AGN (e.g. \citealt{MerloniFabian01}). 

At a redshift of $z=0.184$ \citep{Torres97}, the radio-quiet quasar PDS\,456 has a bolometric 
luminosity of about $10^{47}$\,erg\,s$^{-1}$ and is one of the most luminous nearby AGN \citep{Simpson99, Reeves00}. 
Its black hole mass, estimated from virial scaling relations, is near to $10^{9}$\,M$_{\odot}$ \citep{Reeves09,Nardini15}.  
As a result, the QSO likely accretes close to the Eddington limit. It is known for its ultra fast wind, with an outflow velocity of $0.25-0.3c$, 
which has become well established through many X-ray observations over the last decade \citep{Reeves03,Reeves09,Behar10,Reeves14,Gofford14,Nardini15,Hagino15,Matzeu16,Matzeu17a,Matzeu17b,Parker18,Reeves18a,Reeves18b,BM19,Reeves20}.  

Furthermore PDS\,456 is rapidly variable in X-rays, as was first noticed from X-ray flares detected in {\it RXTE} and {\it Beppo-SAX} observations \citep{Reeves00, Reeves02}, with the X-ray flux increasing by factors of two on time-scales down to tens of kiloseconds and it was noted to be unusual for a high luminosity quasar.  
Subsequently, its rapid variability has also been measured in long {\it Suzaku} observations in 2007 and 2013 \citep{Matzeu16, Matzeu17b}. In particular, the flare seen with {\it Suzaku} in 2013 was the most prominent flare observed to date \citep{Matzeu16}; the X-ray flux increased by a factor of three on a timescale of less than a day, although unfortunately due to scheduling constraints the decline phase was missed.
The 2007 {\it Suzaku} flares were less prominent (up to a factor of two in flux), but were accompanied by significant spectral variability, with the photon index varying between $\Gamma=2.0-2.4$ during the flaring intervals \citep{Matzeu17b}. 
Importantly, the large SMBH mass in PDS\,456, of $M_{\rm BH}\sim10^9$\,M$_{\odot}$ also implies that a gravitational radius translates 
into a lightcrossing time of 5\,ks. This makes it possible to directly probe the compact size-scales close to the SMBH, on observable time-scales, which may not be possible in an AGN of much lower mass. 

Here we present new {\it XMM-Newton}, {\it NuSTAR} and {\it Swift} observations of PDS\,456, 
performed in September 2018. The observations highlight the extremely variable nature of the X-ray corona in PDS\,456. 
The {\it Swift} monitoring reveals a bright X-ray flare, with a total duration of about one week and where the X-ray flux increased by a factor of four. Coordinated {\it XMM-Newton} and {\it NuSTAR} observations, performed $1-3$\,days after the flare peak, capture the decline phase of the flare in detail. 
The paper and analysis is organised as follows. The details of the observational campaign are described in Section~2, along with the {\it Swift} light-curves. 
Section 3 quantifies the broad-band X-ray spectrum of PDS\,456, obtained in a bright state with {\it XMM-Newton} and {\it NuSTAR}, 
following the peak of the flare. Here the hard X-ray cut-off and coronal temperature are measured for the first time in 
PDS\,456, revealing an unusually cool X-ray corona in this epoch. 
In Section~4, the spectral variability during the flare is quantified, where the X-ray emission became substantially harder, 
while Section~5 details the subsequent changes to the optical to hard X-ray Spectral Energy Distribution (SED). Section~6 discusses the physical properties of 
the variable X-ray corona in PDS\,456, which may characterise the high energy emission from an AGN that accretes at a high Eddington rate.
Note that all errors in the text and tables are quoted at 90\% confidence for one interesting parameter (or $\Delta\chi^2=2.7$). A luminosity distance for 
PDS\,456 of $D_{\rm L}=860$\,Mpc was adopted, for cosmological parameters of $H_{\rm 0}=73$\,km\,s$^{-1}$\,Mpc$^{-1}$ and $\Omega_{0}=0.73$. 

\section{Description of Observations}

\subsection{XMM-Newton and NuSTAR Observations}

PDS 456 was observed once with \xmm\ in September 2018 and twice in September 2019, 
as part of a campaign to study the long-term variability of its X-ray ultra fast outflow; see \citet{Reeves20} (hereafter paper I) for a detailed description of all of the observations. 
In this paper, the \xmm\ observation in September 2018 is presented in detail, which was also performed simultaneously with {\it NuSTAR} and alongside a 
concerted {\it Swift} monitoring campaign. 
As is discussed further below, this observation caught PDS\,456 in an exceptionally bright flaring state, in contrast to the low flux 2019 observations which were presented 
in paper I. The details of the 2018 {\it XMM-Newton}, {\it NuSTAR} and {\it Swift} observations are listed in Table~\ref{tab:obslog}. 

\begin{table}
\begin{threeparttable}
\centering
\caption{Observations of PDS 456 in 2018.}
\begin{tabular}{l c c c}
\toprule
Observation & Start (UT)$^{a}$ & Stop (UT)$^{a}$ & Exp(ks)$^{b}$ \\
\midrule
{\it XMM-Newton} & 09/20 14:22:48 & 09/21 13:01:12 & 61.1 \\
{\it NuSTAR} & 09/20 01:16:09 & 09/21 21:46:09 & 81.9 \\
{\it Swift} Obs 1--18 & 08/22 15:52:57 & 09/13 09:16:54 & 46.9 \\
{\it Swift} Obs 19--23 & 09/14 15:16:58 & 09/21 07:03:53 & 12.1 \\
{\it Swift} Obs 24--33 & 09/22 21:07:27 & 10/06 03:53:56 & 28.2 \\
{\it Swift} Obs 18 & 09/13 07:16:45 & 09/13 09:16:54 & 3.1 \\
{\it Swift} Obs 19 & 09/14 15:16:58 & 09/14 17:09:53 & 2.5 \\
{\it Swift} Obs 20 & 09/18 16:34:37 & 09/18 18:23:52 & 2.9 \\
{\it Swift} Obs 21 & 09/19 22:44:01 & 09/19 23:02:52 & 0.9 \\
{\it Swift} Obs 22 & 09/20 03:43:56 & 09/20 05:30:53 & 2.8 \\
{\it Swift} Obs 23 & 09/21 05:01:52 & 09/21 07:03:53 & 2.9 \\
{\it Swift} Obs 24 & 09/22 21:07:27 & 09/22 22:53:53 & 2.9 \\
\bottomrule
\end{tabular}
\begin{tablenotes}
\small
\item $^a$Start and Stop times in MM/DD and HH:MM:SS in UT. 
\item $^{b}$Exposure in ksec with {\it XMM-Newton} EPIC-pn, {\it NuSTAR} FPMA or {\it Swift} XRT. 
\end{tablenotes}
\label{tab:obslog}
\end{threeparttable}
\end{table}

\begin{figure*}
\begin{center}
\rotatebox{270}{\includegraphics[height=16cm]{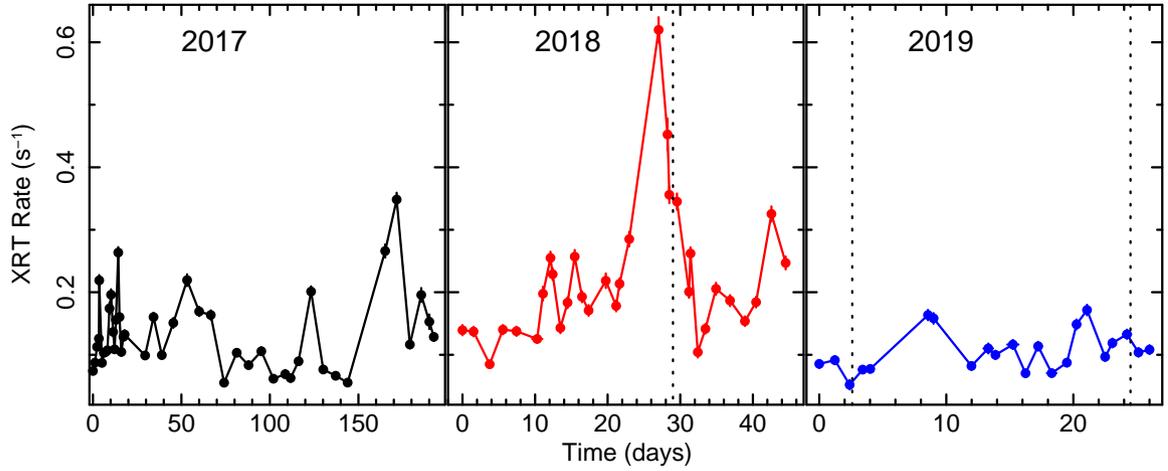}}
\end{center}
\caption{0.3-10\,keV band {\it Swift} XRT lightcurve of PDS\,456 from 2017--2019, extracted from the monitoring campaign and illustrating 3 years of variability of the quasar. 
Note that the time axis corresponds to the time since the start of the monitoring in each year. 
The vertical dotted lines indicate the start times of the three {\it XMM-Newton} observations in 2018 and 2019. In particular the 2018 {\it XMM-Newton} 
caught PDS\,456 in a bright state, approximately 2 days after a major flare in the lightcurve. In contrast the 2019 {\it Swift} XRT monitoring revealed PDS\,456 in a much lower flux, 
occurring at a minimum during the 2019a observation. Note that the count rate errors on some data points are smaller than the marker sizes.}
\label{fig:lightcurve}
\end{figure*}

The observations were processed using the \textsc{nustardas} v1.8.0, 
{\it XMM-Newton} \textsc{sas} v18.0 and \textsc{heasoft} v6.25 software.
{\it NuSTAR} source spectra were extracted using a 50\arcs\ circular region centred on the source and background from a 76\arcs\ circular region clear from stray light. 
{\it XMM-Newton} EPIC-pn spectra were extracted from single and double events in Large Window mode, using a 30\arcs\ source region and $2\times34\arcs$\ background regions on the same chip. 
The spectra and responses from the individual FPMA and FPMB detectors on-board {\it NuSTAR} 
were combined into a single spectrum after they were first checked for consistency.  
The spectra are binned to at least 50 counts per bin and over-sampled the resolution by no more than a factor of three. 
Photometric data-points were also extracted from the 
{\it XMM-Newton} Optical Monitor (OM, \citealt{Mason01}) images, taken with the V, U, UVM2 and UVW2 filters. We corrected for Galactic reddening by adopting the 
extinction law from \citet{Cardelli89} with $R=3.1$ and $E(B-V)=0.45$. 

After background subtraction, the EPIC-pn spectrum resulted in a net count rate of $3.985\pm0.008$\,cts\,s$^{-1}$ over the 0.4--10 keV band and a 
net exposure of 61.1\,ks after correcting for detector deadtime. The background level was very low, $<0.4$\% of the net source rate. 
The pn source spectrum was tested for the possible presence of photon pile-up using the \textsc{sas} task \textsc{epatplot}. 
The ratio of singles to double pixel events were found to be within 1\% of the expected nominal values and thus no significant pile-up is present. 
Spectra from the hard X-ray {\it NuSTAR} FPMA and FPMB detectors were included in the analysis over the 3--50\,keV band, 
with net count rates of $0.207\pm0.002$\,cts\,s$^{-1}$ and $0.200\pm0.002$\,cts\,s$^{-1}$ respectively, while the net exposure per detector was 81.9\,ks and 81.6\,ks 
respectively.  Here the background rate is only 5\% of the source rate, although the spectrum becomes background dominated above 50\,keV where the {\it NuSTAR} effective area declines. 

Spectra from the {\it XMM-Newton} Reflection Grating Spectrometer (RGS, \citealt{denHerder01}) were extracted using the \textsc{rgsproc} pipeline and were combined into a single spectrum, after first checking that the individual 
RGS 1 and RGS 2 spectra were consistent with each other within the errors. The total net count rate obtained over the 6--27\AA\ band was $0.197\pm0.002$\,cts\,s$^{-1}$ 
with a net exposure time of 76.3\,ks per RGS. 

\subsection{Swift Monitoring}

PDS\,456 has also been monitored by the {\it Swift} satellite from 2017--2019. A total number of 32 observations were conducted in 2018 over an approximate 6 week period 
from 2018/08/22 to 2018/10/06 in order to coincide with the visibility window of the {\it XMM-Newton} and {\it NuSTAR} observations (see Table~\ref{tab:obslog}). The observations were performed with a typical 
exposure time of 2-3\,ks from the X-ray Telescope (XRT, \citealt{Burrows05}) and with imaging exposures taken with both the V and UVW1 band filters on-board the {\it Swift} UVOT \citep{Roming05}. 
Monitoring was also performed in 2017 over a longer 200 day baseline from 2017/03/23 to 2017/10/09, consisting of 45 individual exposures, on an approximate daily sampling during the first 18 days and the remainder of the monitoring with a roughly weekly sampling.
Note the very start of the 2017 {\it Swift} campaign 
also coincided with a joint {\it XMM-Newton} and {\it NuSTAR} observation; the observations have been presented in \citet{Reeves18b} and were at a low flux level.  
Monitoring was resumed for about a month from 2019/08/31 to 2019/09/26, consisting of 22 pointings, in order to coincide 
with the two 2019 {\it XMM-Newton} observations published in paper I. 

Figure\,1 shows the complete {\it Swift} XRT lightcurve of PDS\,456 from the whole monitoring campaign, extracted over the 0.3--10\,keV band. 
The three portions correspond to the 2017, 2018 and 2019 monitoring periods as described above. The start times of the three \xmm\ observations in 2018 and 2019 are also marked with vertical dotted lines. Critically, the 2018 monitoring period captured 
a bright X-ray flare centred at T+27 days, where the count rate increased by a factor of 4 and down again over a baseline of just over a week. While flaring is also evident in the 2017 monitoring, these events occur at an overall lower count rate. The 2018 \xmm\ and {\it NuSTAR} observations occurred just two days after the maximum in the {\it Swift} lightcurve and captured part of the decline phase of the flare. 
In contrast, the two \xmm\ observations in 2019 (hereafter 2019a and 2019b) occurred at a much lower flux level, where in particular the {\it Swift} XRT count rate at the time of the 2019a observation was only $\sim0.05$\,cts\,s$^{-1}$, at the minimum flux level of the monitoring and more than an order of magnitude lower than at the peak of the flare in 2018. 
As was shown in paper I, the two low flux 2019 observations revealed a wealth of absorption features from the fast wind from PDS\,456, both in the iron K-shell band and at soft X-rays, while the 2019a observation was also heavily obscured at soft X-ray energies by a low ionization absorber with a column density of $N_{\rm H}\approx10^{23}$\,cm$^{-2}$.  
Figure~1 of \citet{Reeves20} showed the broad-band spectral comparison between these three observations. 

As the main aim in this paper is to study the 2018 X-ray flare, {\it Swift} XRT spectra were extracted from various intervals based upon the 2018 lightcurve. The 32 {\it Swift} observations in 2018 are numbered from obs\,1--33, where observation 15 was not performed due to a GRB trigger. These intervals are listed in Table~1 and are displayed in Figure 2, 
where the latter shows the 2018 XRT lightcurve binned per observation. Here, observations 1--18 occur prior to the flare (the pre-flare period), observations 19--23 coincide with the flare, while observations 24--33 occur in the post-flare period. In addition, XRT spectra from each of the individual 
observations from obs\,18--24 were also extracted to follow the spectral evolution during the flare, with obs\,20 occurring at the peak of the flare. In addition, photometry and fluxes were extracted from each of the V and UVW1 images obtained from the {\it Swift} UVOT, making it possible to monitor changes in the optical to X-ray SED during the flare. 

\begin{figure}
\begin{center}
\rotatebox{-90}{\includegraphics[height=8cm]{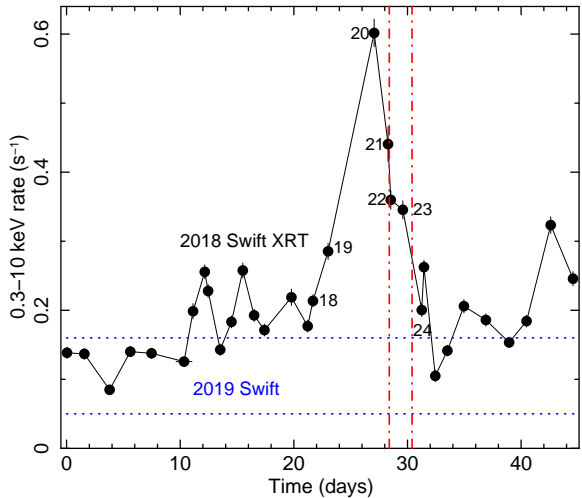}}
\end{center}
\caption{A zoom in of the 0.3--10\,keV {\it Swift} XRT lightcurve during the 2018 monitoring campaign. The time axis corresponds to the start of the 2018 {\it Swift} 
monitoring, as listed in Table\,1. The XRT points are shown as black circles, while the observations 
corresponding to the X-ray flare are numbered from 18--24, with obs\,20 occurring at the peak of the flare. The start and stop times of the {\it NuSTAR} observation are marked by the vertical dot--dashed red lines which occurred during the decline phase of the flare, at about two days after the flare peak. The {\it XMM-Newton} observation was centred 
near to {\it Swift} obs\,23. For comparison the range of count rates observed in the 2019 {\it Swift} monitoring is shown by horizontal blue dotted lines, illustrating the 
quiescent flux level of PDS\,456 captured one year later.}
\label{fig:swiftobs}
\end{figure}


\section{The 2018 {\it XMM-Newton} and {\it NuSTAR} Spectrum}

First the broad-band X-ray spectrum from the joint \xmm\ and {\it NuSTAR} observations was analysed, in order to parametrize the overall spectral form during the 
bright flaring epoch in 2018. The EPIC-pn and NuSTAR spectra were fitted jointly over the 0.4--50\,keV band, where for the latter the combined FPMA and FPMB 
spectra are used, as they are consistent within errors. 
A constant multiplicative factor was also included between the {\it NuSTAR} and {\it XMM-Newton} spectra to allow for any cross normalisation differences between satellites; however, this was found to be consistent with 1.0 within errors. 
The latest version of the \textsc{xspec} \textsc{tbabs} 
model \citep{Wilms00} was used to account for neutral absorption due to our own Galaxy, which includes the fine structure around the O K-shell and Fe L-shell photoelectric edges. Abundances were set to those of \citet{Wilms00}. The Galactic absorption component was subsequently well determined in all of the models below, where $N_{\rm H}=2.57\pm0.04\times10^{21}$\,cm$^{-2}$, which is just slightly in excess of the predicted H\,\textsc{i} column density of $N_{\rm H}=2.4\times10^{21}$\,cm$^{-2}$ based on 21 cm measurements \citep{Kalberla05}. This indicates there is little excess of neutral absorption towards PDS\,456 and is also consistent with the RGS measurements (see Section~3.3).

\begin{figure}
\begin{center}
\rotatebox{-90}{\includegraphics[height=8.5cm]{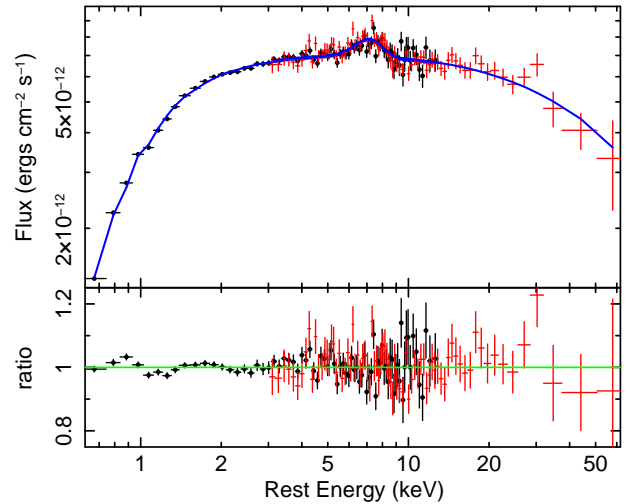}}
\end{center}
\caption{The broad-band {\it XMM-Newton} and {\it NuSTAR} spectrum of PDS 456 in 2018. The model fitted (solid blue line) consists of a powerlaw continuum, absorbed by only the Galactic column of $N_{\rm H}=2.57\pm0.04\times10^{21}$\,cm$^{-2}$, which accounts for the downturn observed at low energies. The continuum is also modified at high energies by an exponential cut-off function with an e-folding energy of $E_{\rm fold}=51^{+11}_{-8}$\,keV, while a 
broad Gaussian emission line near 7 keV was also included. {\it XMM-Newton} EPIC-pn points are shown as black circles and {\it NuSTAR} as red crosses. The lower panel shows the data/model residuals. Spectra are binned by a further factor of $\times3$ for clarity and are plotted in the QSO rest frame.}
\label{fig:cutoff}
\end{figure}

A simple powerlaw with the above Galactic absorption returned a poor fit, with a reduced chi-squared of $\chi_{\nu}^{2}=810/449$ and a roll-over in the {\it NuSTAR} spectrum is apparent at high energies above 10 keV. The continuum form was subsequently modified to include a high energy exponential roll-over (or cut-off powerlaw), 
where the hard X-ray continuum declines in the form of the multiplicative factor $M(E)=\exp[(E_{\rm c}-E)/E_{\rm fold}]$. Here, $E_{\rm fold}$ corresponds to the e-folding energy at high energies, while the low energy cut-off was set to $E_{c}=0.01$\,keV, well below the soft X-ray bandpass. This substantially improves the fit statistic by $\Delta\chi^2=206$ for $\Delta\nu=1$ fewer degrees of freedom. 
Adding a broad Gaussian emission line to parametrize the iron K line emission also significantly improved the fit ($\Delta\chi^2=107$ for $\Delta\nu=3$). 
No soft X-ray excess is required to fit the data below 2\,keV and the whole continuum is well described by the simple cut-off powerlaw. 

\begin{table}
\begin{threeparttable}
\centering
\caption{2018 {\it XMM-Newton} and {\it NuSTAR} Spectral Parameters.}
\begin{tabular}{l c c c}
\toprule
Parameter & Cut-off PL & Reflection & Comptt\\
\midrule
$\Gamma$ & $1.92\pm0.02$ & $1.85\pm0.02$ & -- \\
$E^{a}_{\rm fold}$ or kT$^a$ & $50.7^{+10.7}_{-7.6}$ & $43.5^{+6.4}_{-5.4}$ & $13.1^{+4.4}_{-2.1}$ \\
$\tau^{b}$ & -- & -- & $2.15^{+0.30}_{-0.36}$ \\
$N^{c}_{\rm CPL}$ & $3.57\pm0.03$ & -- & --\\
$N_{\rm H}^{d}$ & $2.57\pm0.04$ & $2.70\pm0.05$ & $2.62\pm0.03$\\
$E_{\rm Gauss}^{e}$ & $6.92\pm0.16$ & -- & $6.84\pm0.19$\\
$\sigma_{\rm Gauss}^{e}$ & $0.83^{+0.19}_{-0.15}$ & -- & $0.91^{+0.23}_{-0.17}$\\
$F^{f}_{\rm Gauss}$ & $2.9^{+0.7}_{-0.6}$ & -- & $3.3^{+0.8}_{-0.7}$\\
EW$^g$ & $230^{+55}_{-48}$ & -- & $257^{+62}_{-54}$\\
$q^{h}$ & -- & $2.2\pm0.4$ & --\\
$\log\xi^i$ & -- & $3.1\pm0.1$ & --\\
$\theta^j$ & -- & $67^{+23}_{-8}$ & --\\
$\mathcal{R}^k$ & -- & $0.14\pm0.04$ & --\\
$F^{l}_{\rm 2-10\,keV}$ & 9.5 & -- & --\\
$L^{m}_{\rm 2-10\,keV}$ & 9.3 & -- & --\\
$L^{m}_{\rm 0.3-50\,keV}$ & 26.7 & -- & --\\
$\chi_{\nu}^{2}$ & 496.1/445 & 494.3/444 & 500.8/445 \\
\bottomrule
\end{tabular}
\begin{tablenotes}
\small
\item$^{a}$E-folding energy of cut-off powerlaw continuum or coronal temperature (comptt model) in units of keV.
\item$^{b}$ Optical depth of corona
\item$^{c}$Normalisation of the cut-off powerlaw continuum in units of $\times10^{-3}$\,photons\,cm$^{-2}$\,s$^{-1}$\,keV$^{-1}$ at 1\,keV.
\item$^{d}$Galactic column density in units of $\times10^{21}$\,cm$^{-2}$.
\item$^{e}$Centroid energy and width of Fe K Gaussian emission line in units of keV.
\item$^{f}$Line flux in units of $\times10^{-5}$\,photons\,cm$^{-2}$\,s$^{-1}$.
\item$^{g}$Equivalent width of emission line in units of eV.
\item$^{h}$Powerlaw disc emissivity.
\item$^{i}$Log ionization parameter of reflector. Units of $\xi$ are ergs\,cm\,s$^{-1}$
\item$^{j}$Inclination in degrees.
\item$^{k}$ Reflection fraction, corresponding the ratio of the reflected flux seen at infinity to the incident continuum flux.
\item$^{l}$Observed 2--10\,keV flux, in units of $\times10^{-12}$\,erg\,cm$^{-2}$\,s$^{-1}$.
\item$^{m}$Absorption corrected rest-frame X-ray luminosity, in units of $\times10^{44}$\,erg\,s$^{-1}$.
\end{tablenotes}
\label{tab:broadband}
\end{threeparttable}
\end{table}

The fit to this model is shown in Figure\,3 and the model parameters are listed in Table~\ref{tab:broadband}. Overall the fit statistic is $\chi_{\nu}^{2}=496/445$ and only low-level residuals are present against the model; applying a small additional systematic error of $\pm1$\% results in an acceptable 
fit statistic of $\chi_{\nu}^{2}=460/445$. The photon index is $\Gamma=1.92\pm0.02$, while the {\it NuSTAR} spectrum is steeper above 10\,keV and this is well 
described by the exponential roll-over with $E_{\rm fold}=50.7^{+10.7}_{-7.6}$\,keV. Only a single broad emission line at iron K is required by the data, with a centroid energy of $6.92\pm0.16$\,keV, a width of $\sigma=0.83^{+0.19}_{-0.15}$\,keV and an equivalent width of $230^{+55}_{-48}$\,eV. The energy of the line profile suggests an origin in He or H-like iron and is similar in terms of its parameters to the broad ionized emission line found in previous observations of this quasar (e.g. \citealt{Reeves09, Nardini15}). 
As per other observations of PDS\,456, there is no requirement for a narrow, neutral Fe K$\alpha$ line at 6.4\,keV, with a tight upper limit on its equivalent width of $<13$\,eV. This is consistent with the X-ray Baldwin effect, whereby the Fe K$\alpha$ equivalent width diminishes with increasing X-ray luminosity and/or Eddington ratio \citep{IwasawaTaniguchi93, Bianchi07}. 

In contrast to previous observations of PDS\,456 (e.g. \citealt{Reeves09, Gofford14, Nardini15, Matzeu17a, Reeves20}) there is no strong requirement for a blue-shifted iron K absorption line associated with a fast wind at this epoch. An upper-limit to its equivalent width of $<65$\,eV was found upon attempting to include a Gaussian line of negative normalisation in the 8--10\,keV range. Indeed the lack of any strong absorption like residuals either in the iron K band or at soft X-ray energies (see Sections 3.3 and 3.4) suggests that only the bare X-ray continuum is observed in this bright 2018 epoch. Here, PDS\,456 may be more similar to some of the bare Seyferts 1s reported in the literature; e.g. Ark\,120 \citep{Vaughan04, Matt14, Porquet18}, Fairall\,9 \citep{Lohfink12, Yaqoob16}, TON\,S180 \citep{ParkerMillerFabian18, Matzeu20}. 

\subsection{Ionized Reflection}

We also tested whether the continuum and broad iron K emission line could be fitted with ionized reflection from an accretion disc. 
The \textsc{relxill} (v1.3.10) reflection model \citep{Garcia14} was used in place of the Gaussian line. This reflection model combines 
the ionized reflection calculations of \citet{Garcia13}, with the relativistic blurring computed with the \textsc{relline} model \citet{Dauser13}, in a self consistent manner. 
The input continuum was assumed to be of the same cut-off powerlaw form.  
The iron line profile, while broad, is not strongly redshifted and as a result the black hole spin was fixed at $a=0$; equivalent fits in terms of $\chi^2$ where 
found between zero ($a=0$) and maximal spin ($a=0.998$). The inner radius was fixed to the innermost stable circular orbit (ISCO), equivalent 
to $R_{\rm in}=6R_{\rm g}$, while the outer disc radius was set to $R_{\rm out}=400R_{\rm g}$. The powerlaw disc emissivity function, in the form of 
$R^{-q}$ was allowed to vary, as were the disc inclination and the ionization and normalisation of the reflector. The reflection fraction ($\mathcal{R}$) was subsequently computed, giving the ratio of the reflected flux observed at infinity to the continuum flux incident upon the disc. 
A Solar abundance of iron was assumed, as otherwise it is not well determined.

The results of the fit are also listed in Table~\ref{tab:broadband}. It is statistically equivalent to the one with a broad Gaussian line, while the continuum 
parameters, in particular the e-folding energy of the cut-off powerlaw, remain very similar. The emissivity index is relatively flat, with 
$q=2.2\pm0.4$, suggesting that the disc is fairly evenly illuminated per unit surface area as a function of radius. 
Note that the inclination angle is constrained to $\theta>59\degg$. The ionization 
parameter is high, with $\log\xi=3.1\pm0.1$, with most iron being in the form of He and H-like ions, as might be expected with the 
centroid energy of the iron line being close to 7\,keV. 
The reflection strength is modest, with $\mathcal{R}=0.14\pm0.04$ and indicates that the spectrum is mainly continuum dominated. 
Overall, the reflector is consistent with originating from an ionized disc, 
where the illumination is not too centrally concentrated, as might be produced by a more extended corona. Indeed, if a lamp-post type geometry is 
adopted for the reflection, then a high upper-limit of $h<19R_{\rm g}$ is found.  

\begin{figure*}
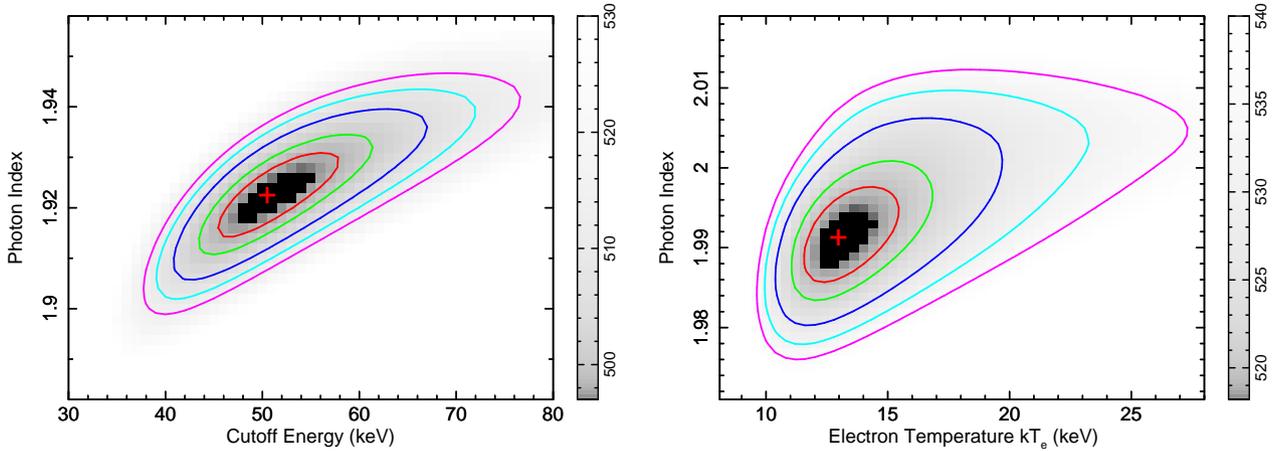

\begin{center}
\rotatebox{-90}{\includegraphics[height=8.5cm]{f4a.eps}}
\rotatebox{-90}{\includegraphics[height=8.5cm]{f4b.eps}}
\end{center}
\caption{Confidence contours for:- (left) the e-folding (cut-off) energy against photon index for a cut-off power-law model and (right) for the electron temperature versus photon index for the thermal Comptonization model, \textsc{nthcomp}. Confidence contours (inner to outer) correspond to the 68\%, 90\%, 99\%, 99.9\% and 99.99\% 
confidence levels for two interesting parameters. The greyscale shading corresponds to the $\chi^2$ values of the fit, as indicated by the right hand bar, with the best-fitting parameter space represented by the darker shades. The cut-off energy and thus the electron temperature are well determined, via the presence of the strong roll-over in the {\it NuSTAR} spectrum towards high energies. Note that the measured cut-off and coronal temperature are at the lower end of what is typically measured recently in other nearby AGN with {\it NuSTAR} (e.g. \citealt{Fabian15,Tortosa18}).}
\label{fig:contour}
\end{figure*}

\subsection{Thermal Comptonization Models} 

Given the presence of the high energy cut-off in the {\it NuSTAR} data, the broad-band spectrum was fitted with thermal Comptonization models. In such models, the seed UV disc photons are inverse Compton scattered by electrons in the X-ray corona with a thermal distribution characterised by a temperature of $kT_{\rm e}$, which could account for the high energy roll-over. The \textsc{nthcomp} \citep{ZdziarskiJohnsonMagdziarz96, ZyckiDoneSmith99}  and \textsc{comptt} \citep{Titarchuk94} 
models were adopted. The results are reported in Table~\ref{tab:broadband} for the latter and these models also give a good description of the 
shape of the X-ray spectrum from PDS\,456 in the 2018 epoch. An input (seed photon) temperature of $kT=10$\,eV was assumed, which is a plausible value for the inner disc blackbody temperature for a luminous quasar accreting at near to Eddington. The broad Gaussian emission line was also retained as before. 

In Figure 4 the confidence contours are shown for the simple cut-off powerlaw model (in the $\Gamma$ vs $E_{\rm fold}$ plane) versus the \textsc{nthcomp} model (in the 
$\Gamma$ vs $kT_{\rm e}$ plane). For each case either the e-folding energy or the coronal temperature are well constrained by the data, where for the latter model $kT_{\rm e}=13.1^{+3.0}_{-1.8}$\,keV.  Similar results were found for the \textsc{comptt} model for either a slab or spherical geometry. For the slab, $kT_{\rm e}=13.1^{+4.4}_{-2.1}$\,keV with an optical depth of $\tau=2.15^{+0.30}_{-0.36}$. For the sphere, $kT=12.9^{+2.7}_{-2.0}$\,keV and $\tau=5.0\pm0.6$. The fit statistic is the same for both cases 
($\chi_{\nu}^2=501/445$) and the higher depth for the sphere is due to the geometry. The values for the slab-like corona are listed in Table~2.  
The coronal temperature is in line with approximate relation with the e-folding energy seen in other AGN, where $E_{\rm fold}\approx2-3kT$ 
(e.g. \citealt{Fabian15}). 

Note that we also checked to see whether the Comptonisation results hold for the case where we use a reflection model to model the iron K line, as in Section~3.1. 
We use the \textsc{relxillCp} variant of the \textsc{relxill} reflection model, which uses the Comptonized continuum computed by \textsc{nthcomp} for the input continuum. The 
coronal temperature is consistent with the above values, where here $kT=14.4^{+1.8}_{-1.7}$\,keV.

The 2018 observation is the first time where it has been possible to measure the high energy cut-off (and thus the coronal temperature) in PDS\,456. This is likely due to a fortunate combination of the high X-ray flux and the fact that the bare continuum has been observed, unmodified by wind absorption unlike in previous {\it NuSTAR} observations of the quasar (e.g. \citealt{Nardini15}). However we note that the measured $kT_{\rm e}$ is significantly lower than what has been typically measured in most AGN with {\it NuSTAR} \citep{Fabian15, Fabian17, Tortosa18, Middei19}. We will discuss further the coronal properties of PDS\,456 in Section~6.  

\begin{figure}
\begin{center}
\rotatebox{-90}{\includegraphics[height=8.5cm]{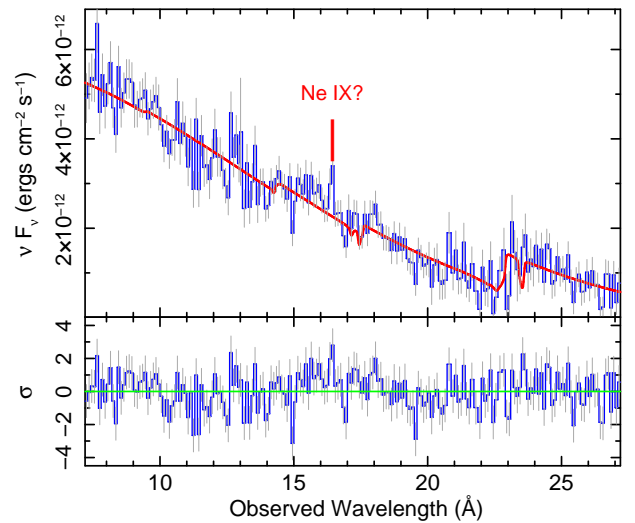}}
\end{center}
\caption{The combined RGS\,1+2 spectrum from the 2018 observation, plotted in the observed frame. The spectrum is binned at $\Delta\lambda=0.1$\,\AA\ resolution and the $1\sigma$ error bars are shown in grey. The solid red-line show a power-law continuum, of $\Gamma=1.98\pm0.08$, absorbed by a neutral Galactic column of $N_{\rm H}=2.8\pm0.3\times10^{21}$\,cm$^{-2}$. Note that the feature near 23\,\AA\ is due to the neutral O\,\textsc{i} edge due to our Galaxy. The lower panel shows the residuals 
(in $\sigma$) against this continuum. The only marginally significant feature is at 16.4\,\AA\ (13.8\,\AA\ rest frame) and may be associated with 
a weak Ne\,\textsc{ix} forbidden emission line (see red marker). Otherwise no significant features are found, in emission or absorption.}
\label{fig:rgs}
\end{figure}

\subsection{The 2018 RGS Spectrum}

The high resolution 2018 RGS spectrum of PDS\,456 was also analysed in order to place constraints on any soft X-ray emitting or absorbing gas in the bright state. 
The RGS spectrum was grouped into bins of width $\Delta\lambda=0.1$\,\AA\ and  $\Delta\lambda=0.05$\,\AA; the former was used to define the continuum and the latter, which slightly over samples the RGS resolution, was used to search for any lines. The continuum was fitted with a simple power-law and neutral Galactic absorption, this yielded 
$\Gamma=1.98\pm0.08$ and $N_{\rm H}=2.8\pm0.3\times10^{21}$\,cm$^{-2}$, consistent with the continuum parameters from the broad-band \xmm\ pn and {\it NuSTAR} spectrum. The RGS spectrum also shows no soft excess, with the same photon index found in the soft X-ray band as at higher energies. 
The O abundance is consistent with that of \citet{Wilms00}, with $A_{\rm O}=1.10\pm0.20$. 
The overall fit statistic is reasonable ($\chi_{\nu}^2=255/207$), which reduces to close to $\chi_{\nu}^2=1.0$ upon the addition of a $\pm3$\% systematic (see \citealt{Kaastra11}) due to the high signal to noise. The observed 0.45--2.0\,keV flux is $4.8\times10^{-12}$\,ergs\,cm$^{-2}$\,s$^{-1}$ (or $9.5\times10^{-12}$\,ergs\,cm$^{-2}$\,s$^{-1}$ when corrected for Galactic absorption).

The RGS spectrum is plotted in Figure\,5, with the residuals to the power-law shown in the lower panel. No strong line-like residuals are present over the spectrum. 
None the less we performed a blind Gaussian search for any line features, adopting the binning at $\Delta\lambda=0.05$\,\AA\ and noting any features (in emission or absorption) 
significant at the $\Delta\chi^2>9.2$ level (or $>99$\% confidence for 2 parameters). The only formally significant feature is an emission line at an observed wavelength of $16.40\pm0.06$\,\AA\ (rest-frame $13.85\pm0.05$\,\AA). This is quite close to the expected wavelength of the Ne\,\textsc{ix} forbidden line (at 13.7\,\AA). The equivalent width is small (${\rm EW}=5.1\pm2.1$\,eV), the line is unresolved ($\sigma_{\rm v}<1200$\,km\,s$^{-1}$) and is significant at the level of $\Delta\chi^2=12.7$. 
No other lines, in emission or absorption, are observed close to their expected positions in the observed frame; e.g. O\,\textsc{vii} He$\alpha$ (forbidden, 26.17\,\AA, or resonance, 25.57\,\AA), O\,\textsc{viii} Ly$\alpha$ (22.46\,\AA) or Ne\,\textsc{x} Ly$\alpha$ (14.37\,\AA). An upper-limit can be placed on the column density of a warm absorber, where the velocity is allowed to vary within $\pm5000$\,km\,s$^{-1}$ of the expected line wavelengths.  For an ionization constrained within the range of $\log\xi=0-2$\,erg\,cm\,s$^{-1}$ and for a turbulence of $\sigma=300$\,km\,s$^{-1}$, then $N_{\rm H}<7\times10^{20}$\,cm$^{-2}$. This is similar to the bare Seyferts, such as Ark 120 (e.g. \citealt{Reeves16}). 


\subsection{Constraints on a Fast Wind in 2018}

In paper I, evidence was found in the low flux 2019b spectrum of PDS\,456 for systematically blue-shifted absorption lines arising from the strong resonance transitions of 
O\,\textsc{vii} Ly$\alpha$, Ne\,\textsc{ix} He$\alpha$ and Ne\,\textsc{x} He$\alpha$. These were found to originate from a blue-shifted absorber, with a column density of 
$N_{\rm H}=2.3^{+0.9}_{-0.6}\times10^{21}$\,cm$^{-2}$, an ionization parameter of $\log\xi\sim3$ and an outflow velocity of $v/c=-0.257\pm0.003$. It was found that 
the outflow velocity was entirely consistent with that inferred for the iron K absorption profile and it was suggested that the soft X-ray absorber may originate from the outer (pc scale) regions of the fast accretion disc wind in PDS\,456. Furthermore in the lowest flux 2019a epoch, the soft wind column increased to $N_{\rm H}\sim10^{23}$\,cm$^{-2}$, although the low soft X-ray flux prohibited a detailed RGS analysis of that epoch. 

We can compare the 2019b RGS spectrum with the bright state 2018 RGS spectrum, where the latter has a three times higher flux over the 0.4--2.0\,keV RGS band. 
The strongest absorption line in the 2019b RGS spectrum originated from 
O\,\textsc{viii} Ly$\alpha$ at a rest wavelength of 14.6\,\AA\ (17.2\,\AA\ observed frame). In the 2018 spectrum we find only an upper limit on the magnitude of the 
equivalent width, of $<3.6$\,eV, significantly weaker than the value of $11.5^{+3.9}_{-3.2}$\,eV measured in the 2019b spectrum. Similar tight upper limits are also found for the blue-shifted Ne\,\textsc{ix} and Ne\,\textsc{x} lines in the 2018 spectrum. This indicates that the strength of the soft X-ray wind has diminished in 2018, possibly as a result of 
the higher continuum flux and subsequent ionization state. 


Constraints were placed on the iron K absorption from a fast wind in 2018, using the \xmm\ pn spectrum. 
The same absorption grid as per the 2019 data in paper I was used, 
which had a turbulence velocity of 10\,000\,km\,s$^{-1}$ to model the strong iron K absorption lines observed in the 2019a and 2019b data-sets.  
This grid was based on the SED of PDS\,456 obtained from the simultaneous 2017 \xmm\ and {\it NuSTAR} observations, which occurred at the start of the 2017 
{\it Swift} monitoring (see \citealt{Reeves18b} for further details). The 2017 SED is appropriate for the 2019 observations, as the X-ray flux was at a similar 
low level. However for the 2018 observation, the X-ray flux was substantially higher, relative to the steady optical/UV flux in PDS\,456 (e.g. \citealt{Hamann18}). 

The effect of this continuum change on the gas ionization state was subsequently investigated. 
To quantify this difference and in particular its effect on the 
iron K band absorber, we compared the 7--30\,keV band fluxes; i.e. directly above the iron K-shell edge threshold. For the 2018 observation, the 7--30\,keV band flux measured with {\it NuSTAR} is $F_{\rm 7-30\,keV}=7.6\times10^{-12}$\,erg\,cm$^{-2}$\,s$^{-1}$, which is about 5 times higher than the flux measured in the 2019b observation 
($F_{\rm 7-30\,keV}=1.5\times10^{-12}$\,erg\,cm$^{-2}$\,s$^{-1}$) or compared to the low flux 2017 observation ($F_{\rm 7-30\,keV}=1.8\times10^{-12}$\,erg\,cm$^{-2}$\,s$^{-1}$). Hence, given the difference in the ionizing X-ray flux, one should expect to observe an increase in the ionization parameter of the absorber by about the same factor if the gas responds in proportion to the hard X-ray continuum.  

To test this, the 2018 spectrum was initially fitted adopting the same column density as per the low flux 2019b observation (where $N_{\rm H}\sim7.0^{+1.9}_{-1.6}\times10^{23}$\,cm$^{-2}$; see Table~3, \citealt{Reeves20}), 
allowing the $N_{\rm H}$ to vary only within the bounds of the errors from the 2019b observation. The velocity was allowed to vary within $\pm0.05c$ of the 2019 values (from $0.2-0.3c$). A lower limit on the ionization parameter for the 2018 observation was found, with $\log\xi>5.75$, which is at least a factor of four higher in linear space 
compared to the 2019b spectrum ($\log\xi=5.02\pm0.12$, \citealt{Reeves20}). This is consistent with the ionization of the wind changing in proportion to the hard X-ray  
flux, for a given column density. Thus the greater hard X-ray luminosity in the 2018 observation might account for the relative weakness of the wind features in that epoch.


\begin{figure}
\begin{center}
\rotatebox{-90}{\includegraphics[height=8.5cm]{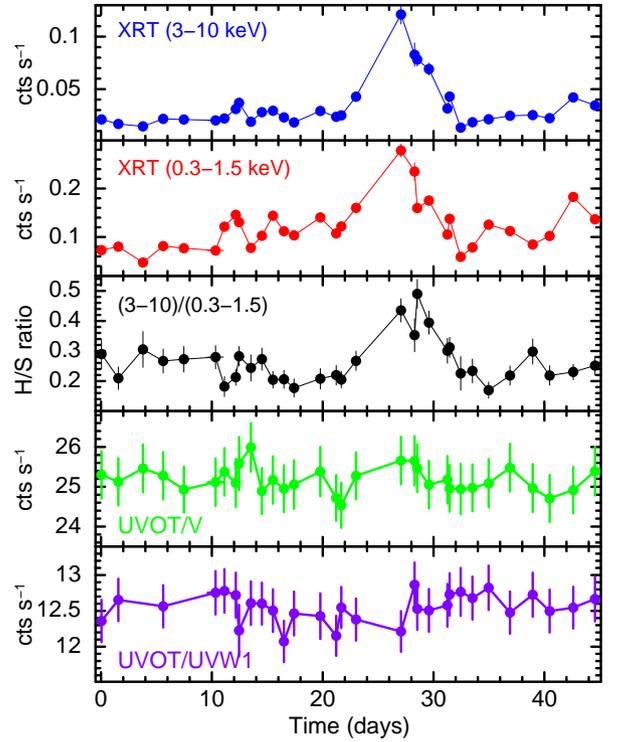}}
\end{center}
\caption{Net-rate multi-band 2018 Swift XRT and UVOT lightcurve of PDS 456. The panels, from top to bottom, show the 3--10\,keV (hard band) and 0.3--1.5\,keV (soft band) XRT lightcurves, the ratio of the hard/soft lightcurves and the Swift UVOT V and UVW1 band lightcurves. There is a clear increase in the source hardness during the course of the flare. 
The slight drop in hardness at $T+28$\,days occurs in obs\,21, which has the shortest exposure of all the {\it Swift} observations.
In contrast, there is no variability in either the optical (V band) or UV (UVW1 band) lightcurves. Note, for some data-points, the marker sizes may be larger than the error bars.}
\label{fig:swiftmulti}
\end{figure}

\section{The X-ray Flare} \label{sec:flare}

We have seen that the 2018 broad band X-ray spectrum of PDS\,456, characterising the decline phase of the flare in Figure~2, can be well described by the thermal Comptonized 
emission from an X-ray corona. It is also devoid of any features from a fast wind, in contrast to the other epochs of this quasar observed from 2001--2019 \citep{Reeves03, Reeves09, Reeves14, Reeves18a, Reeves18b, Reeves20, Behar10, Gofford14, Hagino15, Nardini15, Matzeu16, Matzeu17a, Matzeu17b, Parker18, BM19}. 
Next, the properties of the X-ray flare are described in more detail, while the variability of the X-ray spectrum is also analysed. The latter is important in this context, as it can determine the evolution of the X-ray corona during the flare. 

The multi-band {\it Swift} lightcurve is shown in Figure\,6. Here the XRT lightcurve is split into a hard (3--10\,keV) and soft (0.3--1.5\,keV) band, while the ratio of the hard/soft 
lightcurve is also displayed. In addition the optical (V band) and UV lightcurves (UVW1 band, centred at 2600\,\AA, \citealt{Breeveld11}) are plotted from the {\it Swift}/UVOT. 
Considering the X-ray lightcurves, the magnitude of the flare is greater in the harder band compared to the soft X-rays. This is clear in the hard/soft band ratio, 
which shows that X-ray emission is hardest at the peak of the flare and reveals a gradual decline in the hardness ratio during the decline phase of the flare. 
In complete contrast, no variability is observed in either the V or UVW1 band lightcurves, even during the flare interval. This strongly suggests that the variability of PDS\,456 
is confined to the X-ray corona, with no underlying variability from the accretion disc emission in the optical and UV bands. 

\begin{figure}
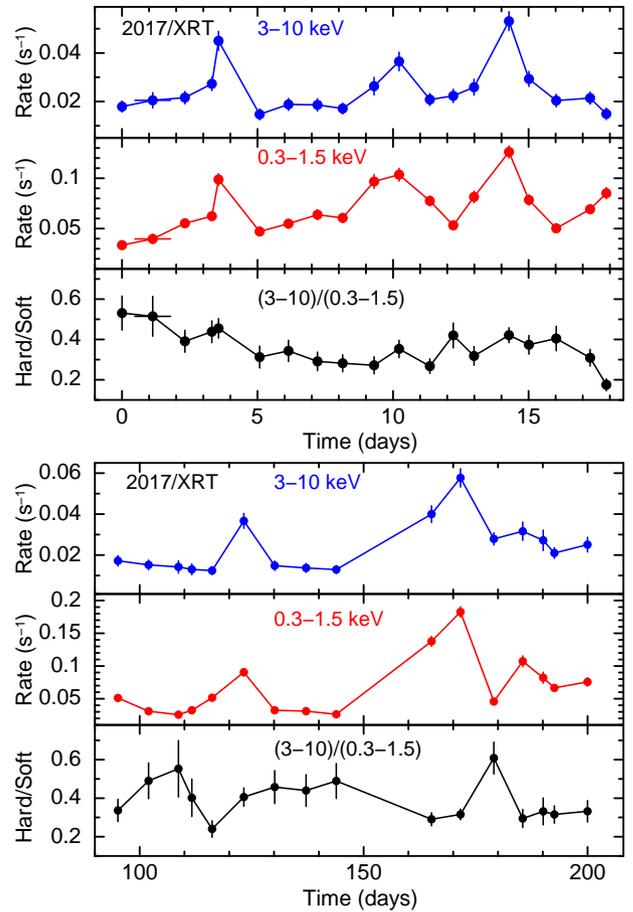

\begin{center}
\rotatebox{-90}{\includegraphics[height=8.5cm]{f7a.eps}}
\rotatebox{-90}{\includegraphics[height=8.5cm]{f7b.eps}}
\end{center}
\caption{As per Figure~6, but showing the hard, soft and hard/soft ratio {\it Swift} XRT lightcurves for a portion of the 2017 monitoring. The upper three panels show the first portion of the lightcurve, with a daily monitoring pattern and where short flaring episodes are seen. The lower three panels show the last 100 days of the monitoring, with a roughly 
weekly sampling and where a prominent flare was observed at $T+170$\,days. In contrast to the 2018 flare, no strong hardness ratio variations accompanied the 2017 flares.}
\label{fig:swift2017}
\end{figure}

For comparison, a similar analysis was also performed for the 2017 {\it Swift} observations, where we also compared the hard band, soft band and hard/soft ratio XRT lightcurves. 
For illustration, in Figure 7, we zoom into two portions of the monitoring; the first 18 days, where {\it Swift} followed a regular daily sampling (upper 3 panels) and the last 100 days during a roughly weekly sampling of the lightcurve (lower 3 panels). 
In the former, minor flaring episodes are seen of shorter duration ($\Delta t\approx 2$\,days), with flux changes of up to a factor of two. 
In contrast to the 2018 flare, no strong hardness variations were seen over the course of these small flares. In the latter portion of the lightcurve, a more pronounced flare is seen at $T+170$\,days, with a factor of three increase in X-ray flux. Only a small spike in hardness is present just after the peak of the flare. However, the less frequent sampling, plus the lack of \xmm\ and {\it NuSTAR} coverage, preclude a more detailed analysis of the 
spectral evolution during this particular flare. Similar to the 2018 monitoring, no significant variability was seen in the V and UVW1 band lightcurves.

\subsection{Spectral Evolution during the Flare}

As was noted in Section~2 (see Table\,1 and Figure\,2), to study the spectral variability further, the {\it Swift} lightcurve was split into three broad intervals for spectral analysis; the pre-flare interval (obs 1--18), the mean flare interval (obs 19--23) and the post-flare interval (obs 24--33). In addition the {\it Swift} spectra were also extracted from the individual {\it Swift} observations from obs 18--24, to study more closely the evolution during the flare. 
All of these {\it Swift} XRT spectra were fitted with a simple power-law continuum modified by the Galactic absorption column measured from the broad-band fit 
in Section\,3. Given the relative short exposures of the {\it Swift} spectra, this was sufficient to achieve a good fit, while a high energy cut-off is not required 
given the lack of bandpass above 10\,keV. 

\begin{table}
\begin{threeparttable}
\centering
\caption{Evolution of PDS 456 during the 2018 {\it Swift} flare.}
\begin{tabular}{l c c c}
\toprule
Observation & $\Gamma$ & $F_{\rm 2-10\,keV}$$^{a}$ & $L_{\rm 2-10\,keV}$$^{b}$ \\
\midrule
Obs 18  & $2.21\pm0.13$ & $3.9\pm0.3$ & $4.0\pm0.3$ \\
Obs 19 & $2.05\pm0.13$ & $6.2\pm0.5$ & $6.1\pm0.5$ \\
Obs 20 & $1.70\pm0.03$ & $18.3\pm1.2$ & $17.0\pm1.1$ \\
Obs 21 & $1.63\pm0.19$ & $16.3\pm2.3$ & $14.6\pm2.1$ \\
Obs 22 & $1.65\pm0.10$ & $11.4\pm1.0$ & $10.5\pm0.9$ \\
Obs 23 & $1.83\pm0.11$ & $8.8\pm0.7$ & $8.4\pm0.7$ \\
Obs 24 & $1.93\pm0.15$ & $5.2\pm0.5$ & $5.0\pm0.5$ \\
Obs 1--18 & $2.16\pm0.03$ & $3.3\pm0.1$ & $3.2\pm0.1$ \\ 
Obs 19--23 & $1.78\pm0.03$ & $11.4\pm0.4$ & $10.7\pm0.4$ \\
Obs 24--33 & $2.09\pm0.04$ & $4.3\pm0.1$ & $4.2\pm0.1$ \\
\bottomrule
\end{tabular}
\begin{tablenotes}
\small
\item$^{a}$2--10\,keV X-ray flux in units of $\times10^{-12}$\,ergs\,cm$^{-2}$\,s$^{-1}$
\item$^{b}$2--10\,keV X-ray luminosity in units of $\times10^{44}$\,ergs\,s$^{-1}$
\end{tablenotes}
\label{tab:swift}
\end{threeparttable}
\end{table}

The resulting $\Gamma$ values and the 2--10\,keV fluxes and luminosites are reported in Table~\ref{tab:swift}. The evolution of both the photon index and X-ray luminosities during the 
flare are also shown in Figure~8. The X-ray spectrum became significantly harder during the flare. In the pre-flare period (obs 1--18) the photon index is $\Gamma=2.16\pm0.03$, which is well within the typical range for PDS\,456; e.g. where $\Gamma=2.0-2.4$, as measured from previous simultaneous \xmm\ and {\it NuSTAR} observations (e.g. \citealt{Nardini15}). At the peak of the flare in obs\,20, the photon index is substantially harder, with 
$\Gamma=1.70\pm0.08$ and then remains hard during obs\,21 and obs\,22, even as the luminosity of the quasar starts to decline. 
This may hint at a short delay, of about a day, of the spectral response to changes in the flare luminosity, which may arise from the physical extent of the corona (see below). After obs\,22, the source hardness declines 
with the X-ray luminosity and by the time of the post-flare interval (obs\,24-33), the photon index has returned to a more typical $\Gamma=2.09\pm0.04$. 

\begin{figure}
\begin{center}
\rotatebox{-90}{\includegraphics[height=8.5cm]{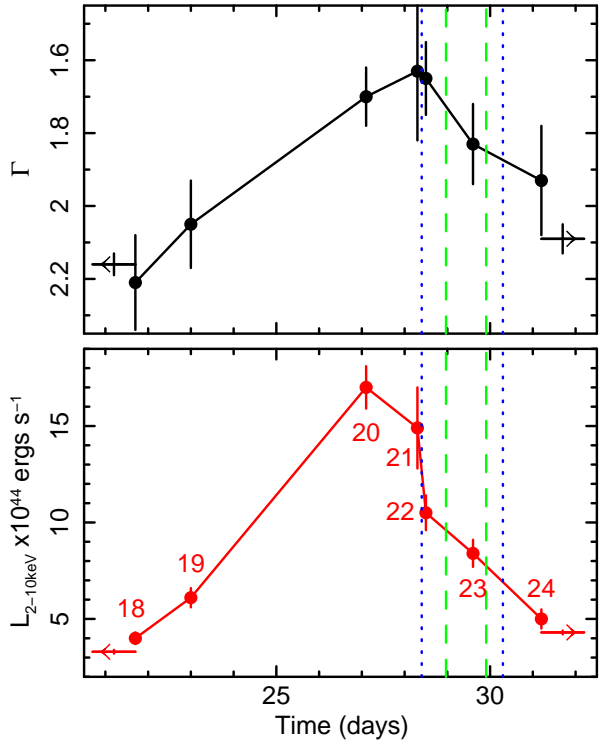}}
\end{center}
\caption{Evolution in the photon index (upper panel) and 2--10\,keV luminosity (lower panel) during the flare. The individual {\it Swift} observations are shown as filled circles as marked, while the summed pre-flare (obs\,1--18) and post flare (obs\,24--33) epochs are indicated by the left and right arrows respectively. The photon index decreases by a factor of $\Delta\Gamma=-0.5$ as the X-ray luminosity increases and reaches its lowest value during obs\,20--22, following the peak of the flare luminosity in obs\,20. 
The photon index steepens after obs\,22, following the decline in the X-ray luminosity. Note that the vertical dotted (blue) horizontal line shows the start and stop time of the 
{\it NuSTAR} and the dashed (green) line the \xmm\ observations. These follow the decline phase of the flare for up to 3 days after the flare peak at T+27 days.}
\label{fig:flare}
\end{figure}

\subsubsection{Fractional Variability}

To quantify further the spectral changes during the {\it Swift} monitoring, we investigate the fractional rms behaviour of PDS\,456 by computing the fractional variability (or $F_{\rm var}$) spectrum. This provides us with a measure of the fractional variability of the source as a function of energy and is a useful tool for exploring source behaviour on different time-scales \citep{Igo20}. Following the description provided in \citet{Vaughan03}, we first extract light curves in a series of adjacent energy bands and then compute the excess variance, $\sigma^{2}_{\rm XS}$, which is defined as: $\sigma^{2}_{\rm XS} = S^{2} - \overline{\sigma^{2}_{\rm err}}$, where $S^{2}$ is the sample variance and $\overline{\sigma^{2}_{\rm err}}$ is the mean square error. We then compute the normalised excess variance by dividing by the squared mean count rate in each band; i.e. $\sigma^{2}_{\rm NXS} = \sigma^{2}_{\rm XS} / \overline{x}^2$. The square root of this value gives the fractional variability, $F_{\rm var}$, which allows us to express the normalised excess variance as a percentage. Errors on $F_{\rm var}$ are given by equation B2 in \citet{Vaughan03}.

This was applied to the {\it Swift} observations over 5 discrete energy bands. Time intervals were chosen which covered; (i) all the 2018 monitoring, (ii) observations 18--24 coincident with the flare, (iii) observations 1--17 and 25--33, which cover the remaining {\it Swift} pointings minus the flare. These are plotted in Figure\,9 and over all the observations it is apparent that $F_{\rm var}$ increases with increasing energy, i.e. greater variability is observed in the harder bands. 
However a striking difference is found from comparing the flare vs non-flare intervals; in the former $F_{\rm var}$ increases with energy, while from the non-flare portion a flat $F_{\rm var}$ spectrum is seen. This is also the case compared to the 2017 $F_{\rm var}$ spectrum, where no enhanced hard X-ray variability is found (Fig 9, blue curve).
This confirms that the 2018 flare is coincident with 
the substantial spectral hardening, as seen from the evolution in $\Gamma$ during the flare, or from the emergence of a hard flaring spectral component. 
In contrast, the spectral hardening is not seen during the 2017 monitoring.

\begin{figure}
\begin{center}
\rotatebox{0}{\includegraphics[height=8cm]{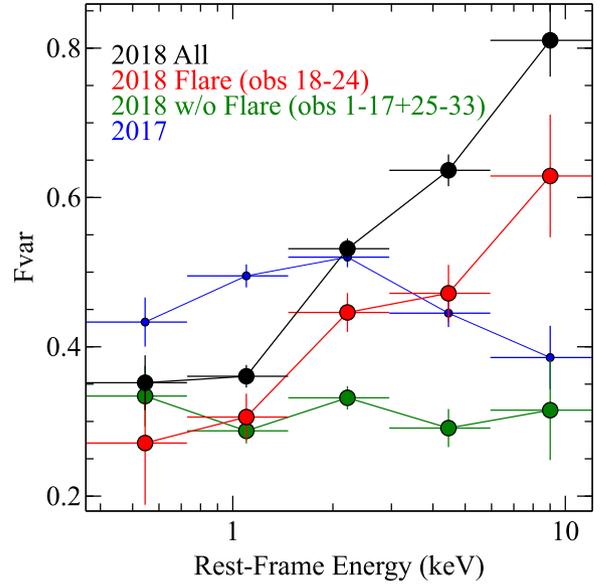}}
\end{center}
\caption{The fractional variability (or $F_{\rm var}$) spectrum of PDS\,456 calculated from the {\it Swift} observations. The black points (upper curve) correspond to all the 2018 {\it Swift} observations, red points (middle curve) are the flare observations (obs\,18--24) and green points (lower curve) are the non-flare periods (obs\,1--17 and obs\,25--33). 
A clear difference is seen in the $F_{\rm var}$ spectrum during the flare and non-flare intervals. During the flare, $F_{\rm var}$ increases strongly with energy, consistent with the variability being driven by the variable hard component. Outside of the flare, the variability is monochromatic, exhibiting no strong spectral evolution. The 2017 $F_{\rm var}$ spectrum is also shown (small blue circles) and does not show enhanced hard variability.}
\label{fig:fvar}
\end{figure}

\subsection{Energetics and Coronal Size}

The total energy of the flare was calculated by integrating the $L_{\rm 2-10\,keV}$ 
curve in Figure~8 between obs\,18--24 and by linear interpolation between data-points. 
The total energy emitted over the 2--10\,keV band is $E_{\rm 2-10 keV}=8\times10^{50}$\,erg. A bolometric correction was defined from the broad-band \xmm\ and {\it NuSTAR} 
spectrum, between the 0.3--50\,keV and the 2-10\,keV bands, which resulted in a correction factor of $\kappa\sim3$. Thus the total energy emitted in X-rays during the flare is $E\sim2\times10^{51}$\,erg\,s$^{-1}$. Note that as there was little variability in the UV and optical bands, this likely represents close to the total output of the flare. 

Between obs\,18--20, the 2--10\,keV luminosity increased by a factor of $4.3\pm0.6$ over a 5 day period. This implies a doubling time for the flare of about 2 days, assuming a linear increase in flux, although the rise time may have been faster if the flare peaked earlier between obs\,19 and 20. On the decline portion of the flare, the halving time is also approximately 2 days. From the light crossing argument, the doubling time implies a X-ray coronal size in PDS\,456 of the order $d\ls c\Delta t \approx 5\times10^{15}$\,cm, which for a black hole mass of $10^{9} {\rm M}_{\odot}$ corresponds to $d\la30R_{\rm g}$. This suggests that the X-ray corona in PDS\,456 
is extended and the flare may originate from the entire X-ray emitting region, rather than just an isolated hot spot or a compact point like source. 

\subsection{The decline phase in detail}

The \xmm\ and {\it NuSTAR} observations were also used to study in greater detail the spectral variability during the declining portion of the flare, due to the better statistics and finer time sampling. 
The {\it NuSTAR} observation occurred over an approximate time-scale of 1--3\,days after the flare peak, commencing just after obs\,21 (and prior to obs\,22) and ending before obs\,24. The shorter \xmm\ observation was centred around {\it Swift} obs\,23; this is visualised by the vertical lines in Figure~8.
The subsequent 3--10\,keV band lightcurve from both \xmm\ (EPIC-pn) and {\it NuSTAR} is shown in Figure~10. 
The lightcurve reveals part of the decaying period, but with more structure than is possible from the {\it Swift} monitoring and may suggest that the decline occurs over a series of step--like decreases. In particular, a rapid factor of $\times 2$ drop in flux over $\Delta t=30$\,ks was observed over the middle portion of the curve. 

\begin{figure}
\begin{center}
\rotatebox{-90}{\includegraphics[height=8.5cm]{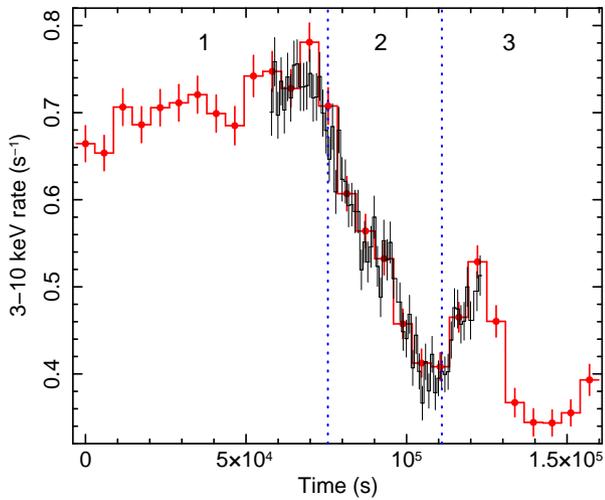}}
\end{center}
\caption{\xmm\ pn and {\it NuSTAR} 3--10\,keV lightcurve during part of the decline phase of the flare, from approximately 1--3 days after the peak. The {\it NuSTAR} points (binned 
to 5814s orbital bins) are shown by the red filled circles, the pn points are in black. In particular, a rapid drop is observed during the \xmm\ part of the curve, centred around {\it Swift} obs\,23. Spectral slices were extracted from the three intervals as indicated, representing (i) a bright plateau period, (ii) a decline phase and (iii) a post-decline period.}
\label{fig:pncurve}
\end{figure}

Three distinct spectral intervals were defined as indicated on Figure~10	:- (i) an initial bright plateau phase, (ii) a decline phase and (iii) a post decline phase. {\it NuSTAR} 
spectra were extracted from each interval, as well as the portions of the \xmm\ pn spectra in each phase. A constant multiplicative factor in flux between the \xmm\ and {\it NuSTAR}  spectra was introduced in the analysis to account for the offset in exposures between intervals (i) and (iii), while the \xmm\ and {\it NuSTAR} intervals were entirely simultaneous in interval (ii) and the constant was set to 1.0. Note that the spectrum in interval~1 does not represent the peak of the flare, as a strong drop in flux was observed 
by {\it Swift} just prior to this interval, occurring between obs\,20 and obs\,22 (Figure~8).

\begin{figure*}
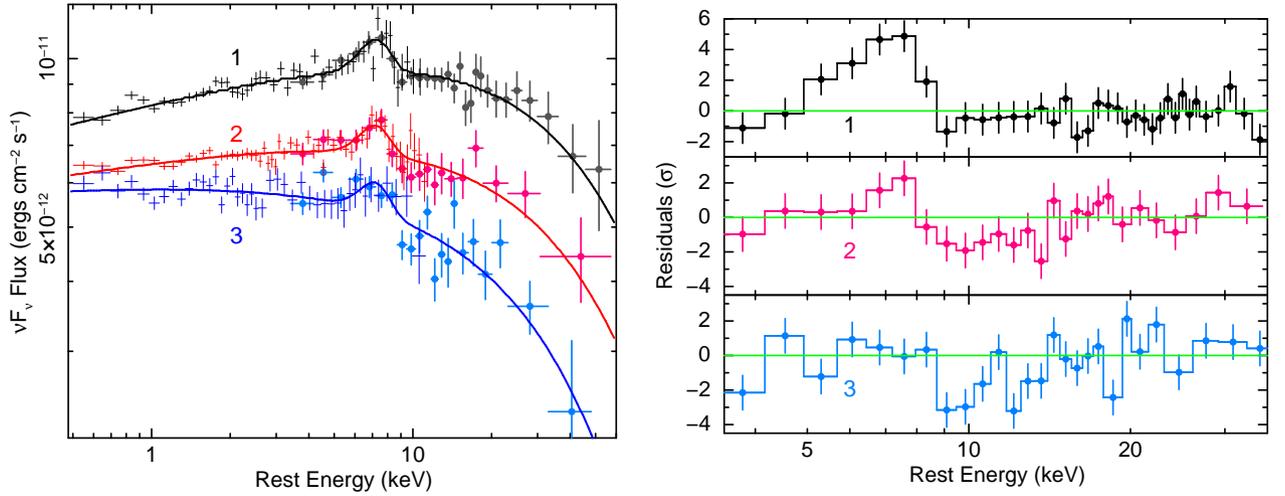

\begin{center}
\rotatebox{-90}{\includegraphics[height=8.5cm]{f11a.eps}}
\rotatebox{-90}{\includegraphics[height=8.5cm]{f11b.eps}}
\end{center}
\caption{{\it Left panel.} The fluxed \xmm\ pn (crosses) and {\it NuSTAR} (filled circles) spectra from slices 1--3. The spectra are fluxed against a $\Gamma=2$ power-law and 
corrected for Galactic absorption. The best fit cut-off powerlaw model to each slice is shown as a solid line, including the broad Gaussian emission at Fe K. 
As the source flux declines, the spectra become noticeably softer, consistent with the trend seen in Figure~8.
{\it Right panel.} The residuals of the three {\it NuSTAR} slices against the cut-off powerlaw continuum, prior to including any iron K emission line. The broad emission line 
peaks near 7 keV and appears strongest during the brightest slice 1 and then diminishing as the hard X-ray continuum declines (see Table~4). Note the final slice 3
shows some absorption residuals between 9--10\,keV, which may indicate that the wind starts to re-emerge once the X-ray flux declines.}
\label{fig:slices}
\end{figure*}

The three spectral slices were fitted with a cut-off powerlaw model modified by Galactic absorption, which is shown in Figure~11 (left panel). This provides a good representation of the broad-band X-ray continuum and the high energy roll-over is also apparent in all three slices, as per the mean spectrum.
The parameters from applying this model to all the slices are listed in Table~\ref{tab:slices}, where a broad Gaussian line is also included. Figure~11 (right panels) shows the residuals of the 
{\it NuSTAR} slices against the continuum, before including the broad line in the model. As is discussed below, the iron emission line is particularly apparent in residuals of the bright first slice, peaking close to 7\,keV in the rest frame.  

Overall, as per the {\it Swift} observations, the continuum appears to become softer as the flux declines, where the photon index increases from slice~1 ($\Gamma=1.87\pm0.01$) to slice~3 ($\Gamma=1.98\pm0.02$). 
There is marginal evidence for the e-folding energy to decrease, where in the bright slice 1, $E_{\rm fold}=65^{+14}_{-7}$\,keV and 
in the faintest slice~3, $E_{\rm fold}=36^{+11}_{-3}$\,keV. The combination of the change in $\Gamma$ and $E_{\rm fold}$ reproduces the spectral softening apparent in Figure~11 from slice 1 through to slice 3. 


\begin{table}
\begin{threeparttable}
\centering
\caption{{\it XMM-Newton} and {\it NuSTAR} Spectral Slices.}
\begin{tabular}{l c c c}
\toprule
Parameter & slice 1 & slice 2 & slice 3 \\
\midrule
$\Gamma$ & $1.87\pm0.01$ & $1.92\pm0.02$ & $1.98\pm0.02$\\
$E^{a}_{\rm fold}$ & $65^{+14}_{-10}$ & $48^{+14}_{-9}$ & $36^{+11}_{-7}$\\
$N^{b}_{\rm CPL}$ & $4.51\pm0.03$ & $3.53\pm0.02$ & $3.13\pm0.04$\\
$E_{\rm Fe}^{c}$ & $6.92\pm0.16$ & $^t$ & $^t$\\
$\sigma_{\rm Gauss}^{c}$ & $0.96^{+0.22}_{-0.17}$ & $^t$ & $^t$ \\
$F^{d}_{\rm Fe}$ & $7.3\pm1.6$ & $4.2\pm2.0$ & $2.9^{+1.2}_{-1.1}$\\
EW$^e$ & $395^{+85}_{-70}$ & $290\pm140$ & $225^{+105}_{-95}$\\
$C^f$ & $0.95\pm0.02$ & 1.0$^f$ & $1.12\pm0.03$ \\
$F^{g}_{\rm 2-10\,keV}$ & 12.7 & 9.2 & 7.4\\
$F^{g}_{\rm 10-50\,keV}$ & 11.2 & 7.1 & 4.4\\
$L^{h}_{\rm 2-10\,keV}$ & 12.4 & 9.1 & 7.4 \\
$L^{h}_{\rm 10-50\,keV}$ & 11.0 & 7.1 & 4.5 \\
Exp ({\it XMM})$^i$ &16.8 & 36.4 & 8.1\\
Exp ({\it NuSTAR})$^i$ & 38.0 & 17.8 & 22.8\\
$\chi_{\nu}^{2}$ & 396.9/382 & -- & -- \\
\bottomrule
\end{tabular}
\begin{tablenotes}
\small
\item$^{a}$E-folding energy of the cut-off powerlaw continuum in units of keV.
\item$^{b}$Normalisation of the cut-off powerlaw, where the units are $\times10^{-3}$\,photons\,cm$^{-2}$\,s$^{-1}$\,keV$^{-1}$ at 1\,keV.
\item$^{c}$Centroid energy and width of Fe K emission line in units of keV.
\item$^{d}$Line flux in units of $\times10^{-5}$\,photons\,cm$^{-2}$\,s$^{-1}$.
\item$^{e}$Equivalent width of emission line in units of eV.
\item$^{f}$Multiplicative constant between the \xmm\ and {\it NuSTAR} spectra. Fixed at 1.0 for slice 2.
\item$^{g}$Observed flux, in units of $\times10^{-12}$\,erg\,cm$^{-2}$\,s$^{-1}$.
\item$^{h}$Absorption corrected rest-frame X-ray luminosity, in units of $\times10^{44}$\,erg\,s$^{-1}$.
\item$^{i}$Exposure time of slice in ksec from \xmm\ or {\it NuSTAR}.
\item$^{t}$Parameter is tied between the slices.
\end{tablenotes}
\label{tab:slices}
\end{threeparttable}
\end{table}


Curiously, as is seen from Table~\ref{tab:slices}, the flux of the broad iron K line near 7 keV decreases with the declining continuum flux, while as a result its equivalent width remains 
constant (within errors). The iron line flux is about 2.5 times higher in the brightest slice~1 ($F_{\rm Fe}=7.3\pm1.6\times10^{-5}$\,photons\,cm$^{-2}$\,s$^{-1}$), versus the faintest slice~3, where $F_{\rm Fe}=2.9\pm1.2\times10^{-5}$\,photons\,cm$^{-2}$\,s$^{-1}$, while the hard X-ray (10--50\,keV) flux decreases by the same factor. Thus the line emitter may respond to the decline in the continuum. In this context, the relatively large line flux in slice~1 might be due to a response to an earlier high continuum level (e.g. at the flare peak in obs\,20), via a subsequent reverberation delay. 
Unfortunately the short {\it Swift} observations cannot measure the iron line at early epochs to confirm this. None the less, the observed response of the line flux between slice 1 and 3 on a time-scale of within a day corresponds to a light crossing distance of $R\sim3\times10^{15}$\,cm, or $\sim20R_{\rm g}$ for $M_{\rm BH}=10^{9}$\,M$_{\odot}$. Thus it may be that the iron K emission from the inner disc regions is being revealed in this unusually bright, bare state of PDS\,456. 
Note, if we instead parametrize the iron K emission with ionized reflection, as per Section~3.1, then the flux of the reflection component also diminishes with 
decreasing luminosity. 

Finally we also placed limits on the presence of the fast, ionized wind in the three slices. We parameterise it through a simple Gaussian line of negative normalisation. 
In slice 3, there is some evidence (e.g. Figure~11, right, lower panel) of negative residuals between 9--10\,keV in the {\it NuSTAR} data. 
Upon modeling these with a Gaussian line, the equivalent width of the absorption line is $170\pm95$\,eV in slice 3, compared to $<90$\,eV in slice 1 where no residuals are apparent. However, as this change is not significant at the 90\% level, we are unable to further quantify the change of the wind in response to the continuum on these short timescales. 


\subsubsection{Fractional Variability and Difference Spectrum}

The $F_{\rm var}$ spectrum was also computed for the 2018 {\it XMM-Newton} EPIC-pn and {\it NuSTAR} observations. 
For the pn, we do this over 30 approximately equally logarithmically spaced energy bins from 0.3--10\,keV, while for the {\it NuSTAR} spectrum, we use 10 bins from 3--50\,keV. For both data-sets, we extracted light curves with a bin size of $\Delta t = 5\,814$\,s, corresponding to the {\it NuSTAR} orbital time-scale. The 
subsequent $F_{\rm var}$ spectrum is shown in Figure~12 (left panel). Note that the y-axes are slightly different for the two spectra. This is because the {\it NuSTAR} observation covered a longer baseline and so we are sampling variability on a slightly longer time-scale than with {\it XMM-Newton}.
A high minus low difference spectrum was also extracted for comparison and is also plotted in Figure~12 (right panel). In this case, the slice 1 minus slice 2 spectrum was computed, as the slice 3 spectrum has a much shorter \xmm\ exposure. 

\begin{figure*}
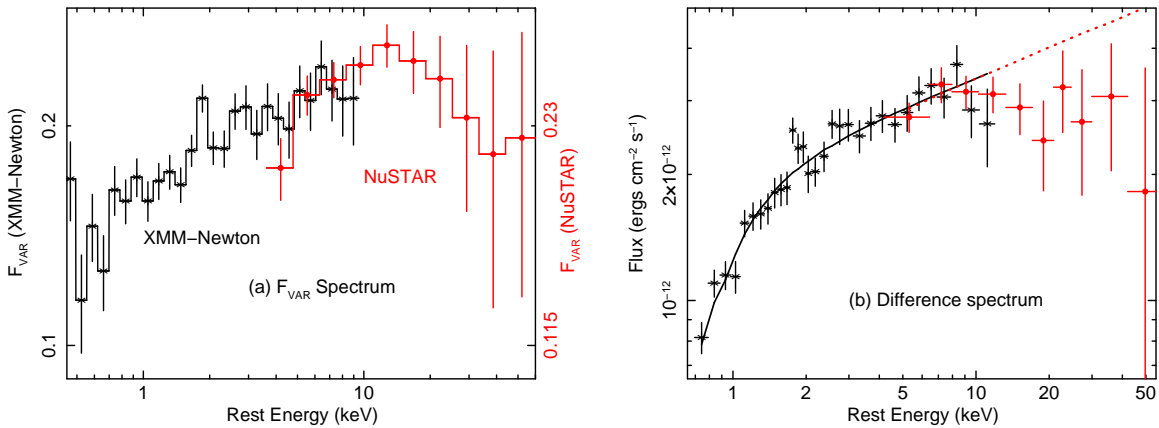

\begin{center}
\rotatebox{-90}{\includegraphics[width=5.5cm]{f12a.eps}}
\rotatebox{-90}{\includegraphics[width=5.5cm]{f12b.eps}}
\end{center}
\caption{(a) The $F_{\rm var}$ spectrum and (b) the difference spectrum, computed from the {\it XMM-Newton} and {\it NuSTAR} observations. Both show a rising 
spectrum below 10\,keV, which is consistent with the variable emission arising from a harder component associated with the flare, with a photon index of $\Gamma=1.7$ (solid black line in the difference spectrum). Above 10\,keV, the spectra harden as the variability starts to diminish towards higher energies. This may be due to the presence of the 
high energy cut-off, which is apparent in all of the hard X-ray spectral slices.}
\label{fig:difference}
\end{figure*}

Both the $F_{\rm var}$ and difference spectra show a hard, rising spectrum below 10 keV from the {\it XMM-Newton} data, indicating that the variability is dominated by a harder spectral component. However in the {\it NuSTAR} band above 10\,keV, the spectra roll over, showing that the variability diminishes towards higher energies.
The difference spectrum has been compared to a power-law of $\Gamma\sim1.7$ fitted below 10 keV, which, when extrapolated above 10\,keV  
over-predicts the variable emission that is seen. Instead the spectrum can be simply parametrized by a broken power-law function, where the photon index steepens from $\Gamma=1.74\pm0.05$ to $\Gamma=2.15\pm0.15$ around a rest-frame break energy of 10\,keV. 
This could also be accounted for by a cut-off power-law, where in this case, given the larger errors, the e-folding energy is constrained at the 90\% level to be between 25--100\,keV. Interestingly, the photon index below 10\,keV is consistent with that obtained in the {\it Swift} spectra near the peak of the flare and thus could be indicative of a harder, Comptonized emission component, which may dominate the flare emission.

The $F_{\rm var}$ and difference spectrum of the flare here is quantitatively similar to that measured during the 2007 {\it Suzaku} observation of PDS\,456, when the QSO was also in a fairly bright, continuum dominated state; see \citet{Matzeu17b} for details. There, a mixture of hard and soft X-ray flares were observed. Although the level of variability in 2007 was more modest, the $F_{\rm var}$ spectrum of the 
{\it Suzaku} hard flares showed a very similar form to that observed here (e.g. compare with Figure~12 and 13 of \citealt{Matzeu17b}), with a rising spectrum up to 10\,keV, consistent with a harder $\Gamma=1.7-1.8$ powerlaw. In contrast, the soft X-ray flares seen appeared to show the opposite behaviour, 
with an increase in the soft X-ray excess (which is not observed here) and a steepening of the 2--10\,keV power-law component up to $\Gamma=2.4$. 
This may suggest a bi-modal behaviour of the X-ray flares seen towards PDS\,456.

\section{The Evolution of the SED during the Flare}

Next, the evolution of the SED of PDS\,456 during the course of the flare is modelled and the results are interpreted in the framework of accretion disc corona models. 
Four different epochs were considered:- (i) the pre-flare SED (formed from the obs\,1--18 {\it Swift} observations), (ii) the peak flare SED ({\it Swift} obs\,20), 
(iii) the decline phase (from the simultaneous {\it XMM-Newton} and {\it NuSTAR} spectrum) and (iv) the post-flare SED ({\it Swift} obs\,24--33). Using the mean {\it XMM-Newton} and {\it NuSTAR} observation for epoch (iii) allows a high signal to noise parameterisation out to hard X-rays of the declining flare phase. As there was no variability in either the optical or UV bands throughout the 2018 campaign, the mean \xmm\ OM and {\it Swift} UVOT photometric points were used to characterise this part of the SED and these are applied to each of the epochs with a 2\% systematic error in flux\footnote{see https://xmmweb.esac.esa.int/docs/documents/CAL-SRN-0378-1-1.pdf for an estimate of the \xmm\ OM systematics}.
The SED of these four epochs are shown in Figure~13. For plotting purposes only, the X-ray spectra have been corrected for Galactic absorption, 
while the optical and UV fluxes have been corrected for reddening, as per Section~2. 

The frequently used \textsc{optxagnf} model \citep{Done12} was then used to describe each of the epochs, which fits the SED from an accreting SMBH in an energetically consistent manner \citep{Jin12}.  
This model comprises of three different emission components which are illustrated schematically in Figure~5 of \citet{Done12}. 
At radii greater than $R_{\rm cor}$, the gravitational potential energy is released in the form of a colour corrected disc blackbody spectrum and this accounts for the optical and UV continuum. Within the coronal radius, $R_{\rm cor}$, the power is radiated in a hot Comptonized component to produce the hard X-ray power-law, with the remainder in the form of a warm Comptonized component and can explain the soft X-ray excess which is often observed in AGN when its temperature is above about 0.1\,keV. 
The fraction of the accretion power below $R_{\rm cor}$ that is 
radiated in the form of the hard X-ray power-law is given by the parameter, $f_{\rm PL}$. 

Here, our aim is to quantify the relative changes of the hard power-law indices and their corresponding $f_{\rm PL}$ values between the epochs, as well as to estimate the scale of the coronal radius and the overall luminosity. 
We note that an in-depth SED spectral analysis, including relativistic and inclination angle effects as performed 
for Ark\,120 \citep{Porquet19}, is beyond the aim of the present paper and will be reported in forthcoming work. 
We used a modified version of {\sc optxagnf} that allows the hot corona temperature to be a free parameter, rather than fixed to 100\,keV (C.\ Done, private communication). 
As no significant soft X-ray excess is observed in the 2018 PDS\,456 spectra, 
the electron temperature of the warm optically-thick component was set to $kT_{\rm e}=0.1$\,keV. 
Thus in the present case, the warm Comptonization component has no observable impact on the observed X-ray spectra, although it may enhance the 
Wien tail of the disc emission in the unobservable extreme UV band. 
A co-moving proper distance of 726\,Mpc\footnote{The luminosity distance, $D_{\rm L}$, is then calculated from the co-moving distance, $D_{\rm CM}$, as $D_{\rm L} = (1+z) D_{\rm CM}$} and a black hole mass for PDS\,456 of $M_{\rm BH}=1\times10^{9} {\rm M}_{\odot}$ were adopted as model input parameters. 
This is consistent with the black hole mass estimates from either \citet{Reeves09} or 
\citet{Nardini15}, computed on the basis of the correlation between H$\beta$ FWHM and 5100\,\AA\ luminosity arising from a virialized BLR; e.g. from \citet{Nardini15}, then 
$\log(M_{\rm BH}/M_{\odot})=9.2\pm0.2$. The effect of adjusting the black hole mass is discussed below. 
The dimensionless black hole spin parameter ($a$) was fixed to $a=0$, and the default inclination of $\cos i=0.5$ is assumed, which is also consistent with the reflection modelling in Section~3.1.

Only two parameters are required to vary between epochs. These are the coronal parameters, the hard X-ray photon index ($\Gamma$) and $f_{\rm PL}$, the fraction of the accretion power radiated in the hot corona. 
 Note that while both the Eddington fraction and the coronal radius are free parameters in the model, they are not allowed to vary {\it between} the four epochs. This is due to the lack of variability in the optical and UV band, as a significant adjustment in either parameter between epochs produces strong variations in the accretion disc emission (e.g. as per the SED of Ark\,120, \citealt{Porquet19}) which is not observed in the {\it Swift} UVOT lightcurve of PDS\,456.  
The results from applying the \textsc{optxagnf} model to the four SED epochs is shown in Table~\ref{tab:SED}. 
The photon index shows the same hardening as observed in the {\it Swift} monitoring, decreasing by $\Delta\Gamma=-0.5$ at the peak of the flare, while $f_{\rm PL}$ increases from $\sim9$\% (pre-flare) to $\sim17$\% (peak flare) and back down again post-flare.
Thus the SED changes observed in Figure~13 are purely described by the variability of the X-ray corona, with no changes in the overall disc emission or accretion rate.  
The best-fitting Eddington ratio is $\log(L/L_{\rm Edd})=-0.38^{+0.06}_{-0.07}$  
for an assumed black hole mass of $M_{\rm BH}=10^{9}$\,M$_{\odot}$, i.e. $\sim 40$\% of Eddington. The total bolometric luminosity is 
$L_{\rm bol}\sim5\times10^{46}$\,ergs\,s$^{-1}$. The coronal radius, which describes the transition between the optically thick disc and the coronal region, 
is $R_{\rm cor}=35^{+8}_{-4} R_{\rm g}$ (or $R_{\rm cor}\sim5\times10^{15}$\,cm). The coronal extent is broadly consistent with the size-scale derived from the flare variability in Section~4. The inferred hot corona temperature of $kT=13^{+3}_{-2}$\,keV is very similar to that found in section~3.2.

The effect of adjusting the black hole mass in the SED model was also investigated. This just results in changes in the Eddington ratio and the coronal radius, 
the other parameters ($\Gamma$, $f_{\rm PL}$) are virtually unchanged. Adopting a smaller mass of $M_{\rm BH}=5\times10^{8}$\,M$_{\odot}$, the 
Eddington ratio is correspondingly higher with $\log(L/L_{\rm Edd})=0.21\pm0.04$ and is mildly super-Eddington. 
Black hole mass values much smaller than this 
are unlikely, as PDS\,456 would then be significantly super-Eddington. The coronal radius is moderately 
larger in terms of the gravitational radius in this case ($R_{\rm cor}=47^{+3}_{-4} R_{\rm g}$), with 
a fit statistic  identical to the model reported in Table~\ref{tab:SED} for a $10^{9}$\,M$_{\odot}$ black hole.

\begin{figure}
\begin{center}
\rotatebox{-90}{\includegraphics[height=8.5cm]{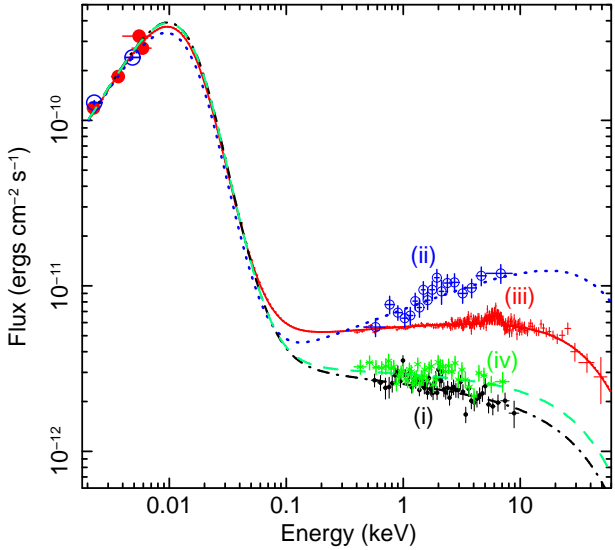}}
\end{center}
\caption{The 2018 optical to X-ray SED of PDS\,456 and fitted by the \textsc{optxagnf} model (see text). The X-ray data-sets correspond to:- 
(i) the pre-flare {\it Swift} spectrum (obs\,1--18, black filled circles), (ii) the peak flare spectrum ({\it Swift} obs\,20, blue open circles), (iii) the mean \xmm\ and {\it NuSTAR} spectrum 
(red crosses, representing the flare decline epoch) and (iv) the post flare spectrum ({\it Swift} obs\,24-33, green stars). The optical and UV photometric points are 
from the {\it Swift} UVOT (blue open circles) and \xmm\ OM (red circles); these have been applied to all epochs as there is no optical or UV variability. 
The SED changes can be purely described by variability in the hot coronal emission, with no resulting change in the overall mass accretion rate.}
\label{fig:sed}
\end{figure}

\begin{table*}
\begin{threeparttable}
\centering
\caption{Multi Epoch SED Model}
\begin{tabular}{l c c c c}
\toprule
Parameter & Decline & Pre-flare & Peak flare & Post-flare \\
\midrule
$M_{\rm BH}/10^{9} {\rm M}_{\odot}$ & 1.0 & -- & -- & -- \\
$\log (L/L_{\rm Edd})$$^{a}$ & $-0.38^{+0.06}_{-0.07}$ & $^t$ & $^t$ & $^t$ \\
$R_{\rm cor}/R_{\rm g}$$^b$ &  $35.0^{+7.5}_{-3.8}$ & $^t$ & $^t$ & $^t$ \\
$kT_{\rm hot}$ (keV) &  $13.1^{+2.8}_{-1.8}$ & $^t$ & $^t$ & $^t$ \\
$\Gamma$ & $1.99\pm0.01$ & $2.27\pm0.04$ & $1.75^{+0.08}_{-0.07}$ & $2.16\pm0.04$\\
$F_{\rm PL}$$^c$ (in \%) & $12.5\pm1.1$ & $8.9^{+1.6}_{-1.3}$& $17.4\pm1.9$ & $8.1^{+1.3}_{-1.1}$ \\
$F^{d}_{\rm 2-10\,keV}$ & $9.4\pm0.8$ & $3.1^{+0.5}_{-0.4}$ & $17.5^{+1.8}_{-1.9}$ & $4.1^{+0.6}_{-0.5}$\\
$F^{d}_{\rm 10-50\,keV}$ & $7.2^{+0.7}_{-0.6}$ & $1.3\pm0.2$ & $22.9^{+2.4}_{-2.5}$ & $2.2\pm0.3$ \\
$\chi_{\nu}^{2}$ & 956/889 & -- & -- & -- \\
\bottomrule
\end{tabular}
\begin{tablenotes}
\small
\item$^{a}$Log ratio of $L_{\rm bol}$ to $L_{\rm Edd}$.
\item$^{b}$Coronal radius, $R_{\rm cor}$, in units of the gravitational radius.
\item$^{c}$Fraction of accretion power below $R_{\rm cor}$ emitted in the X-ray power-law.
\item$^{d}$Observed flux, in units of $\times10^{-12}$\,erg\,cm$^{-2}$\,s$^{-1}$.
\item$^{t}$Parameter is tied between the epochs.
\end{tablenotes}
\label{tab:SED}
\end{threeparttable}
\end{table*}

\section{Discussion} \label{sec:discussion}

\subsection{The Cool X-ray Corona of PDS 456}

For the first time it has been possible to measure the high energy cut-off in the hard X-ray spectrum of PDS\,456, thanks to the high X-ray flux and continuum dominated spectrum from the 2018 {\it NuSTAR} and {\it XMM-Newton} observations. PDS\,456 also appears one of the highest luminosity AGN for 
which the cut-off has been measured. 
The cut-off energy (of $E_{\rm fold}\sim50$\,keV) and thus the coronal temperature (of $kT\sim15$\,keV) revealed for PDS\,456 are substantially lower than measured at hard X-rays for most other AGN, where typical rollover energies of 100\,keV or higher are observed; e.g. with {\it Beppo-SAX} \citep{Perola02, Dadina07}, {\it Swift-BAT} \citep{Ricci18}, {\it Integral} \citep{Malizia14} or {\it NuSTAR} \citep{Fabian15, Tortosa18, Kamraj18}. For instance \citet{Kamraj18} studied the cut-off energies obtained from {\it NuSTAR} observations of hard X-ray selected {\it Swift} BAT AGN. 
Of these, only 6 out of 46 AGN had lower-limits on the cut-off consistent with values of about 50\,keV 
or lower. 

Subsequently, \citet{Middei19} performed a detailed comparison between AGN cut-offs measured with {\it NuSTAR} \citep{Tortosa18} and the predictions made with the {\it MoCA} model (Monte-carlo code for Comptonization in Astrophysics; \citealt{Tamborra18}). The latter performs up to date photon ray tracing simulations of the corona, predicting output Comptonized spectra for a variety of different physical conditions and coronal geometries. 
Overall, 26 AGN in this sample have detected cut-offs in their {\it NuSTAR} spectra, with e-folding energies ranging from $42\pm3$\,keV (Ark 564) to $720^{+130}_{-190}$\,keV (NGC\,5506) when modelled with a cut-off powerlaw. 
By comparison with the predictions from the {\it MoCA} code, \citet{Middei19} computed a corresponding range of coronal temperature 
from $kT=21\pm2$\,keV to $123^{+9}_{-15}$\,keV (for a slab geometry), with mean coronal values of $<kT>=50\pm21$\,keV and $<\tau>1.9\pm0.8$. 

We used the empirical relations derived by these authors between the cut-off energy, photon index, versus the coronal temperature and optical depth (see their equations 2--5 and Figure 6), to deduce the coronal parameters of PDS\,456 from the {\it MoCA} simulations. 
Adopting $E_{\rm fold}=50$\,keV and $\Gamma=1.9$ from the mean 2018 {\it NuSTAR} and {\it XMM-Newton} spectrum, values of $kT\approx 20-25$\,keV and 
$\tau\approx 3$ are derived for PDS\,456. The same temperature was derived for a spherical corona, but with a higher depth of $\tau\sim5$. 
This confirms the unusually low temperature of the hard X-ray corona in PDS\,456, as inferred from a self consistent physical model. 

Indeed, the sensitive imaging hard X-ray observations afforded by {\it NuSTAR} have now revealed a small subset of AGN, which also have 
low cut-off energies. \citet{Tortosa17} first reported an unusually low high energy cut-off in {\it NuSTAR} observations of GRS\,1734$-$292, an AGN which lies close to the Galactic plane. Here a temperature of $kT\sim12$\,keV was measured, with a thick coronal depth (with $\tau\sim3$ for a slab and $\tau\sim6$ for a spherical coronal geometry), similar to the case of PDS\,456.  
As noted above, the NLS1 Ark\,564 has a very similar high energy roll-over \citep{Kara17}, with a measured cut-off energy of $E_{\rm fold}=42\pm3$\,keV and 
a subsequent coronal temperature of $kT=15\pm2$\,keV. Another example is the nearby, luminous QSO, 1H\,0419$-$577, 
where \citet{Turner18} also reported a temperature of $kT=15$\,keV for the X-ray corona as measured its {\it NuSTAR} spectrum. 
This AGN also likely accretes at a high rate and like PDS\,456, has a steep UV to X-ray SED, dominated by a strong blue-bump. 
One possibility is that the softer seed photon population of these AGN helps to cool the corona, resulting in a softer hard X-ray spectrum, 
as was also suggested to account for very soft, X-ray spectrum of the NLS1, RE\,J1034+396 \citep{PoundsDoneOsborne95}. None the less, these steep spectrum hard X-ray AGN appear to be unusual examples compared to the AGN population as a whole and it is possible that a high Eddington ratio may be the link between them. 
This may support the trend observed by \citet{Ricci18}, whereby higher Eddington AGN have cooler hard X-ray corona, based on their {\it Swift}--BAT results. 

\subsection{The Coronal Variability}

The daily {\it Swift} monitoring has made it possible to capture a luminous X-ray flare from PDS\,456, with a total energy output exceeding $10^{51}$\,erg. 
As is seen in both Sections 4 and 5, the flare is only observed in the X-ray band and no significant variability is found in the optical or UV bands. As a result the flare appears intrinsic to the X-ray corona, while the underlying accretion disc emission, which produces the seed photons, remains constant. 
Furthermore, during the flaring periods, PDS\,456 exhibits a pronounced hardening of the spectrum, as seen through changes in the hardness ratio, the $F_{\rm var}$ 
spectrum of the flare and subsequently through the changes in photon index. This is opposite of the behaviour in many radio-quiet AGN, whereby the X-ray spectra 
can become softer with increasing flux and as a result tend to exhibit $F_{\rm var}$ spectra which decline towards higher energies. Recent examples of AGN with steep (or soft) 
$F_{\rm var}$ spectra include:- PG\,1211+143 \citep{Lobban16}, IRAS\,13224-3809 \citep{Parker17}, Ark 120 \citep{Lobban18} and NGC 3227 \citep{Lobban20}. 

Such rapid X-ray variability in PDS\,456 has also been detected in past monitoring observations, e.g. with {\it RXTE} and {\it Beppo-SAX} \citep{Reeves00, Reeves02} and 
was noted as being unusual for a luminous, high mass, radio-quiet quasar. 
\citet{Reeves02} interpreted the earlier X-ray flares seen towards PDS\,456 in terms of a cascade of magnetic reconnection events from the accretion disc corona, with a predicted total output reaching as high as $10^{51}$\,erg, in agreement with the value observed here. 
Similar arguments were also formulated by \citet{MerloniFabian01} in order to explain the X-ray variability of AGN generally.  

The flare time-scale has also made it possible to estimate the likely coronal size. 
The light crossing distance, as derived from the doubling time during the rise of the flare, implies a size of $D\ls5\times10^{15}$\,cm, which 
corresponds to $D\la30 R_{\rm g}$ for a black hole mass of $M_{\rm BH}=10^{9}$\,M$_{\odot}$. The coronal radius deduced from the SED modelling, via the 
\textsc{optxagnf} model (Section~5.1), is also consistent with this. 
Furthermore, from behaviour of the flare in Figure\,8, we note that the response (increase) in the photon index may appear to be delayed by 1--2\,days following the peak of the flare. This could arise as a result of the physical extent of the corona, in order for changes in its physical properties (like the electron temperature) to be 
transmitted throughout the X-ray emission region. 


Changes in the coronal properties, in particular its temperature ($\theta$) and depth ($\tau$), following an injection of magnetic energy into the corona, could reproduce the pronounced spectral changes during the flare.  
Both affect the Compton $y$ parameter in the same direction, where $y=4(\theta + 4\theta^2)\tau(\tau+1)$ (and $\theta=kT/m_{\rm e}c^2$); here 
$kT$ determines the energy transferred per photon scattering and $\tau$ sets the average number of scatterings per photon. 
Either an increase in $\theta$ or $\tau$ will translate into a harder spectrum and an increase in observed X-ray luminosity. 
Figure~6 in \citet{Middei19} also shows how values in the $kT$ versus $\tau$ plane in X-ray coronae 
map on to the observed $\Gamma$ and $E_{\rm fold}$ for the hard X-ray continuum, as computed from the {\it MoCA} code. 
Comparing their results for a slab geometry, the evolution from $\Gamma=2.2$ to $\Gamma=1.7$ (and back again) 
during the flare could be explained by a doubling in 
temperature from $kT=20$\,keV to $kT=40$\,keV, for a constant depth of $\tau\sim3$. The Compton $y$ parameter increases from $y\sim2$ to $y\sim5$ 
as a result. This would explain the harder flare spectrum, with the approximate e-folding energy also increasing (from $\sim50$ to $\sim100$\,keV), although the {\it Swift} XRT spectra are not sensitive to the cut-off. 
Alternatively an increase in optical depth, e.g. via an injection of pairs into the corona during the flare onset, would also result in similar spectral hardening. 

The reverse may happen during the decline phase and the corona could cool due to inverse Compton scattering, if less energy is injected post-flare. 
The post-flare {\it NuSTAR} spectral slices show a noticeable softening of the hard X-ray spectrum following the flare and this 
might suggest a cooling corona. This can occur on plausible time-scales, for instance \citet{Matzeu17b} estimate a Compton cooling time-scale 
of the order of a few tens of kiloseconds in PDS\,456, to account for spectral softening following flares seen in {\it Suzaku} observations. 
Regardless, the cooling time-scales are almost certainly shorter than the light crossing time of the corona, as is discussed further below.

\subsection{Coronal compactness versus temperature}

The measurement of both the temperature and size of the corona in PDS\,456 makes it possible to place the QSO on a diagram of compactness versus temperature; e.g. as recently compiled by \citet{Fabian15} for AGN with a variety of hard X-ray cut-off measurements. 
The compactness parameter is approximately the ratio between the light crossing versus cooling time of the corona 
and is defined as (e.g. \citealt{GuilbertFabianRees83}):-

\begin{equation}
l = \frac{L}{R} \frac{\sigma_{\rm T}}{m_{\rm e}c^3}
\end{equation}

\noindent where $L$ is the luminosity and $R$ the source radius. High compactness values (where $l>>1$) imply Compton cooling times much shorter than 
the light crossing time and electrons will readily cool as they propagate through the corona. For PDS\,456, we use a coronal size of $5\times10^{15}$\,cm, 
while from the SED in Section~5.1 we calculate a UV to X-ray luminosity (from 5\,eV to 100\,keV) of $L_{\rm UV-X}\sim5\times10^{46}$\,erg\,s$^{-1}$.  
This results in a compactness value of $l\sim200$ for PDS\,456. Adopting a lower luminosity will result in a lower compactness; e.g. 
if only the 0.1--200\,keV X-ray band is considered, as per \citet{Fabian15}, then the compactness is an order of magnitude lower, given that the UV luminosity dominates over the X-ray one for PDS\,456. 

Figure\,14 shows the position of PDS\,456 on the $kT$ vs compactness ($l$) plane. The position of most of the AGN compiled in the \citet{Fabian15} 
coronal study is marked by the shaded area. In comparison the dashed line shows the electron-positron pair annihilation  
line, as calculated by \citet{Stern95} for a slab corona (with similar relations for other geometries) and subsequently adopted in the AGN compilations of \citet{Fabian15, Fabian17}. To the right (or above) this line, at a high temperature and compactness, runaway pair production can occur and it is noticeable that most 
AGN with coronal measurements fall in the area just to the left (or below) of the pair annihilation line. As discussed by \citet{Fabian15}, it seems likely that this process 
acts as a thermostat for the corona, an increase in pairs subsequently leading to a reduction in observed luminosity and temperature. 

\begin{figure}
\begin{center}
\rotatebox{0}{\includegraphics[width=8.5cm]{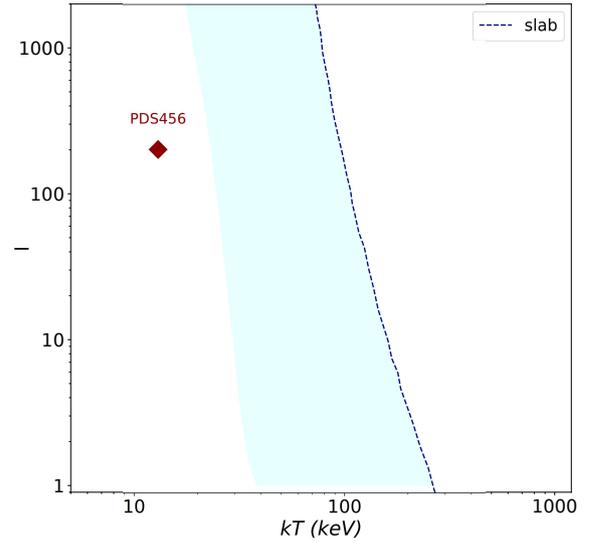}}
\end{center}
\vspace*{-0.5cm}
\caption{The position of PDS\,456 on the temperature ($kT$) vs compactness ($l$) plane. The dashed blue line shows e$^+$/e$^-$ annihilation line for a corona with a slab geometry, adapted from \citet{Fabian15}, above which runaway pair production would occur. The light blue shaded area shows the position of the 
AGN compiled by \citet{Fabian15}, which tend to fall just to the left of the annihilation line, with typical temperatures in the range from 50--200\,keV. On the other hand, PDS\,456 (diamond point) falls well to the left of the diagram, indicating that it has an unusually cool corona compared to most AGN.}
\label{fig:compactness}
\end{figure}

PDS\,456 falls about an order of magnitude below this line in terms of its unusually low temperature, as would other AGN with cool coronae such as 
Ark\,564 and 1H\,0419-577. The coronal measurement of PDS\,456 corresponds to just one point in time, when the corona appears to be in a cooling phase. 
Thus an interesting question is how might PDS\,456 evolve on this diagram during the course of the flare? To fully test this, multi-epoch broad-band observations (i.e. with \xmm\ and {\it NuSTAR}) would be required to quantify how the coronal temperature varies, via the high energy roll-over. 
Thus if the temperature increases during the flare onset as the luminosity rises (and $\Gamma$ hardens), the AGN should move towards the upper-right, closer towards the pair annihilation line and the shaded area in the diagram. This would be as a result of both the temperature and compactness (via the luminosity) increasing. 
The opposite may happen during the decline (cooling) phase.

Another question is how is the coronal temperature regulated and indeed 
how pairs can be produced given the lack of sufficient hard X-ray photons for such a cool corona? 
One possibility is that a significant non-thermal tail exists to the electron population, 
with the Comptonization arising from a hybrid thermal/non-thermal plasma (e.g. \citealt{ZLM93}). Indeed, \citet{Fabian17} discuss this possibility 
in detail and conclude that some of the coolest coronae are likely to have the most substantial non-thermal components. In this case, increasing the 
fraction of non-thermal versus thermal leptons in the corona has the effect of shifting the pair line towards lower temperatures (e.g. compare with Figure\,3, \citealt{Fabian17}), as the pairs are readily produced by the non-thermal tail. This scenario could be compatible for AGN with cool corona like PDS\,456. 

\begin{figure*}
\begin{center}
\rotatebox{0}{\includegraphics[width=16cm]{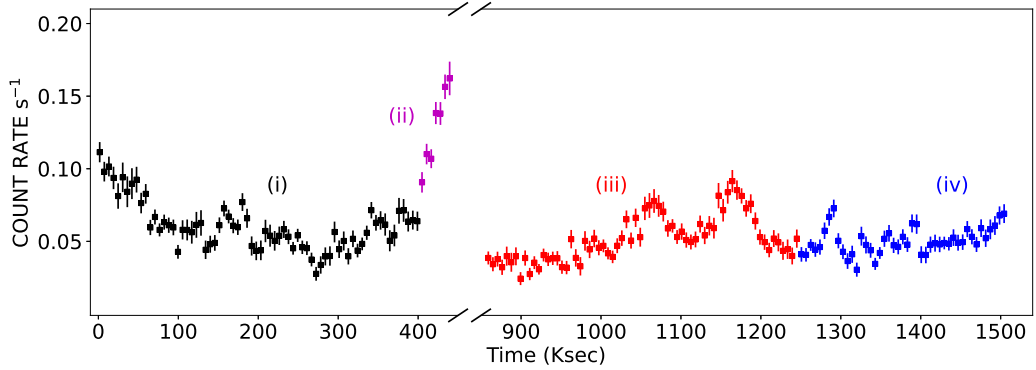}}
\end{center}
\caption{The 2013 {\it Suzaku} XIS lightcurve of PDS\,456, extracted over the 3--10\,keV band. Data from the XIS\,0 and XIS\,3 CCDs were combined into a single lightcurve. 
A $\times3$ increase in flux due to the onset of a pronounced flare is observed 400--450\,ks into the observations. Following a gap due to scheduling constraints, PDS\,456 returned to a low, quiescent flux state after the flare. The observation was split 
into 4 segments (colour coded (i) to (iv), black, magenta, red and blue) for spectral analysis. }
\vspace{-0.5cm}
\label{fig:suzakulc}
\end{figure*}

An alternative non-thermal source of X-ray emission in PDS\,456 may be the base of a relativistic jet. Although PDS\,456 is classified as a radio-quiet quasar \citep{Simpson99, Reeves00}, it is not radio-silent \citep{Yun04}.
One possibility is that the jet activity is sporadic and associated with the strong X-ray flare. 
The X-ray photon index of the variable flare component, of $\Gamma=1.7$, is consistent with the typical values observed in radio-loud quasars \citep{ReevesTurner00, Donato01} and is considerably harder (by $\Delta\Gamma=-0.5$) compared to what is usually observed towards PDS\,456 
\citep{Nardini15}. However not all the previous X-ray flares seen towards PDS\,456 show this behaviour. The lower flux flares seen during the 2017 {\it Swift} monitoring (Figure 1) do not exhibit the same 
change in hardness as per the 2018 flare, while the 2007 {\it Suzaku} observations showed a mixture of hard and soft flares \citep{Matzeu17b}. 

Interestingly, recent VLBI observations of PDS\,456 show evidence for a compact, relativistic jet \citep{Yang19, Yang20}. The high resolution 
radio map of \citet{Yang20} 
shows a collimated jet structure, which extends up to 500\,pc to the NW of the optical nucleus of PDS\,456. Further, non collimated, extended radio structure is also observed, which the authors suggest may be associated to thermal shocks produced by the wide angle wind of PDS\,456. 
However, the jet power is low ($<10^{40}$\,ergs\,s$^{-1}$), more than five orders of magnitude 
below the X-ray luminosity of PDS\,456 and also by the same factor compared to the radio emission typically observed towards flat-spectrum radio-loud quasars of similar bolometric luminosity
(e.g. Figure 10, \citealt{Donato01}). Producing a significant contribution towards the X-ray emission from a low power jet in PDS\,456 may then be challenging. 
None the less, future radio vs X-ray monitoring could establish any link between the onset of radio activity and the strong X-ray 
flares seen towards PDS\,456. 

\begin{figure}
\begin{center}
\rotatebox{-90}{\includegraphics[height=8.5cm]{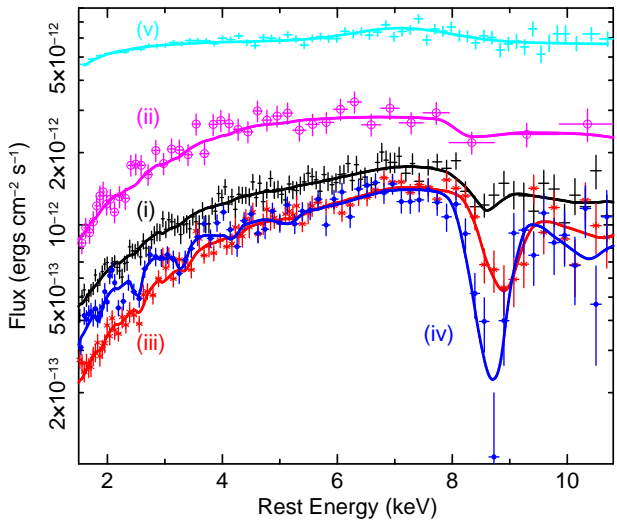}}
\end{center}
\caption{Time selected spectra of PDS\,456 from the 2013 {\it Suzaku} observations. As is indicated on Figure~12, these correspond to; (i, black crosses) the pre-flare spectrum, (ii, magenta, open circles) the flare spectrum, (iii, red stars) the initial post-flare spectrum, (iv, blue, filled circles), the final post flare spectrum. The spectra show a substantial increase in opacity of the 
blue-shifted iron K absorption line between 8.5--9\,keV, occurring in the post-flare segments (iii) and (iv). This can be modeled by an increase in the 
wind column density from $N_{\rm H}\sim10^{23}$\,cm$^{-2}$ prior to the flare, to $N_{\rm H}\sim10^{24}$\,cm$^{-2}$ post-flare. 
For comparison, the mean 2018 \xmm\ spectrum is also shown (v, cyan crosses), which is devoid of features. 
Note the counts spectra have been converted into $\nu F_{\nu}$ fluxed spectra against a $\Gamma=2$ powerlaw and the best fit \textsc{xstar} absorption models (solid lines) are superimposed. }
\label{fig:suzakuspec}
\end{figure}

\subsection{The Connection between X-ray Flares and the Wind}

The bright 2018 observations of PDS\,456 have revealed a featureless X-ray continuum, devoid of strong 
wind features. 
\citet{Matzeu17a} compared the X-ray luminosities for all of the archival \xmm, {\it Suzaku} and {\it NuSTAR} observations of PDS\,456 from 2001--2014; the QSO was found to vary in the range $L_{\rm 2-10\,keV}=2.8-10.5\times10^{44}$\,erg\,s$^{-1}$. 
In comparison, during the 2018 {\it XMM-Newton} and {\it NuSTAR} observation, the X-ray luminosity varied between $L_{\rm 2-10\,keV}=7.4-12.4\times10^{44}$\,erg\,s$^{-1}$ 
(Table~\ref{tab:slices}), at the uppermost end of the historic range, while the luminosity at the flare peak in {\it Swift} 
was even higher, at $L_{\rm 2-10\,keV}=1.7\times10^{45}$\,erg\,s$^{-1}$ (Table~\ref{tab:swift}). 
In contrast, the 2--10\,keV luminosity of the 2019 observations, where strong wind features were revealed, were nearly an order of magnitude fainter ($L_{\rm 2-10\,keV}=2.5\times10^{44}$\,erg\,s$^{-1}$; \citealt{Reeves20}). 
Thus in 2018 the wind features may be weak due to the strong ionizing flux, occurring as a result of the flare. The harder than usual continuum (reaching $\Gamma=1.7$ near the flare peak) could also contribute towards the wind becoming over ionized. 
In future work, we will investigate further whether the intrinsic difference between the iron K profiles between the 2018 and 2019 epochs 
can be accounted for via the continuum changes acting upon self consistent disc wind models \citep{Sim08, Sim10}. 


\subsubsection{Comparison to the 2013 Suzaku Observations}

The comparison with a long 2013 {\it Suzaku} observation of PDS\,456, covering a baseline of over 1\,Ms and sampling sufficient variability, provides an opportunity to test the wind variability in response to a major flare. These data were originally studied in \citet{Gofford14} and \citet{Matzeu16}; here 
we revisited the lightcurve and time-selected spectra to further investigate the wind variability in response to continuum in the light of the above. 

The 3--10\,keV lightcurve of the 2013 {\it Suzaku} observation is shown in Figure\,15. The onset of a substantial flare is seen between 400--450\,ks, with the 3--10\,keV flux increasing by a factor of $\times3$ during this period. Unfortunately, due to a scheduling gap (between 450--850\,ks), the full coverage of the flare was missed. 
Subsequently, in the latter part of the lightcurve, the flux returned to a quiescent level, with only minor flaring present. 
As a result, the XIS (0+3) spectra were extracted from four distinct segments:- (i) the pre-flare portion (0--400\,ks), (ii) the flare onset (400--450\,ks), (iii) an initial post-flare period
(850--1200\,ks) and (iv) a final post-flare period (1200--1500\,ks). To fit the spectra, a baseline model consisting of a powerlaw modified by neutral partial covering absorption (in addition to the Galactic column) was adopted, as is described in detail by \citet{Gofford14} and \citet{Matzeu16}. 
To account for the wind absorption, the  
\textsc{xstar} grid adopted for the low flux 2017 SED of PDS\,456 was used, as was also was recently applied to the 2019 low flux observations \citep{Reeves20}. 
Note the SED of the 2013 {\it Suzaku} observations (Fig~5, \citealt{Matzeu16}) was very similar to the 2017 epoch (Fig~10, \citealt{Reeves18b}), with similar 2--10\,keV fluxes and little optical/UV variability. 

The resulting spectra and best-fitting models are plotted in Figure~16. It is apparent that substantial wind variability is observed, as is seen through the growth of the strong Fe\,\textsc{xxvi} Ly$\alpha$ absorption line through segments (i) to (iii) and (iv) between 8.5--9.0\,keV in the QSO rest frame. 
This can be accounted for by an order of magnitude increase in the wind column density, where in the pre-flare segment (i) a modest $N_{\rm H}=1.4\pm0.9\times10^{23}$\,cm$^{-2}$ is measured, which increases significantly post flare to a maximum of $N_{\rm H}=1.4\pm0.3\times10^{24}$\,cm$^{-2}$ in 
segment (iv). In this case a common (i.e. constant) ionization was assumed, where $\log\xi=5.1\pm0.2$. Alternatively, an order of magnitude decrease in the 
ionization (for a constant column) could also model the increase in opacity (see \citealt{Gofford14, Matzeu16} for details). However such a strong decrease in ionization is not consistent with the continuum variability, where the post-flare flux is only $\sim30$\% lower than the pre-flare level.  Furthermore, the increase in flux during the flare would only serve to initially increase the wind ionization, whereas the opposite change occurs. Thus simple ionization changes imposed upon a steady state wind appear a less likely scenario and the wind may be intrinsically variable in column density in this case. 

\subsubsection{Does coronal activity drive the wind in PDS 456?}

Curiously, the ten-fold increase in opacity in 2013 occurred a few days after the initial flare. This might suggest a link between major X-ray flares, such as in 2013 or 2018 and subsequent wind ejection events. One possibility is that the enhanced radiation flux associated with the flares could help drive the wind ejecta. 
The peak flare X-ray luminosity in 2013 reaches $L\sim3\times10^{45}$\,erg\,s$^{-1}$ (extrapolated over 0.3--50\,keV and absorption corrected). 
The corresponding radiation momentum rate is then $\dot{p}_{\rm rad}\sim10^{35}$\,dyne. For the 2013 wind parameters ($N_{\rm H}=10^{24}$\,cm$^{-2}$, $v=0.25c$) and 
adopting a launch radius equal to the escape radius ($R_{\rm esc}=32R_{\rm g}\sim5\times10^{15}$\,cm), then the mass outflow rate is 
$\dot{M}=\mu m_{\rm p} \Omega v_{\rm out} N_{\rm H} R \sim 5\times10^{26}$\,g\,s$^{-1}$. The latter estimate is consistent with that of \citet{Nardini15}, for 
a derived wind solid angle of $\Omega=2\pi$ in PDS\,456. The corresponding wind momentum rate is then $\dot{p}_{\rm w}=4\times10^{36}$\,dyne. Thus even allowing for an underestimate of the flare luminosity (as the flare peak was not observed in 2013), its radiation thrust is unlikely to be sufficient to drive the emerging wind without a substantial force multiplier factor. The latter appears unlikely for a high ionization wind (e.g. \citealt{KraemerTombesiBottorff18}). 

More plausibly, the wind may be linked to coronal events, leading to the ejection of material through MHD processes (e.g. \citealt{Fukumura15}). 
While this cannot be directly tested on the 2018 observations due to the lack of sensitive X-ray coverage post-flare, the 2013 observations do require the emergence of powerful wind ejecta a few days after the initial flare. The dynamical time-scale appears plausible, where $t_{\rm dyn}=v\Delta t\sim 3\times10^{15}$\,cm; i.e. 
the crossing times are of the order of tens of gravitational radii, commensurate with the inner disc. Curiously, the coronal size inferred from modelling the 2018 flare is also of the order of the escape radius of the $0.25c$ wind component. 
Possible coronal ejection events have been suggested to coincide with strong flares seen in the highly variable Seyfert 1, Mrk 335 \citep{Wilkins15}, where an ultra fast outflow may be observed in higher flux states of this AGN \citep{Gallo19}. Early {\it XMM-Newton} observations of the NLS1, Mrk 766 \citep{Pounds03}, also implied a putative link between flares and disc ejecta, with an increase in iron K absorption being associated with periods of flaring activity. Future monitoring of PDS\,456 will help capture further flares and sensitive follow-up observations will yield insight into the correspondence of the wind with the coronal flares. 

\section{Acknowledgements}

We dedicate this paper to our friend and colleague, Ian George, who has sadly passed away. 
JR acknowledges support through grants 80NSSC18K1603 and HST-GO-14477. VB acknowledges support from grant 80NSSC20K0793. GM and APL both acknowledge the support of an ESA fellowship. DP acknowledges financial contribution from CNES.
EN acknowledges financial contribution from ASI-INAF n.2017-14-H.0. We also thank Chris Done for use of the modified \textsc{optxagnf} model. 
Based on observations obtained with XMM-Newton, an ESA science mission with instruments and contributions directly funded by ESA Member States and NASA. 
This work has been partially supported by the ASI-INAF program I/004/11/4

\section{Data Availability}

The data analysed in this paper are publicly available from the {\it XMM-Newton} (http://nxsa.esac.esa.int/) and 
\textsc{NASA/HEASARC} (https://heasarc.gsfc.nasa.gov) archives. 



\bibliographystyle{mnras}
\bibliography{biblio} 


\bsp	
\label{lastpage}
\end{document}